\documentclass[a4paper, 12pt]{article}

\usepackage{amsmath}
\usepackage{mathtools}
\usepackage{amssymb}
\usepackage{amsthm}
\usepackage{todonotes}
\usepackage{setspace}
\usepackage{fullpage}
\usepackage{placeins}
\usepackage{minibox}
\usepackage{graphicx}
\usepackage{mathtools}
\usepackage{tikz}
\usepackage{xcolor}
\usepackage[capitalise, noabbrev]{cleveref}
\usepackage{caption}
\usepackage{booktabs}
\usepackage{multirow}
\usepackage{authblk}
\usepackage{epstopdf}
\usepackage{algorithm}
\usepackage{algpseudocode}
\usepackage{circledsteps}
\usepackage{soul}
\usepackage{comment}
\usepackage{url}

\usepackage{bm}

\crefname{algorithm}{Algorithm}{Algorithms}

\newtheorem{proposition}{Proposition}

\DeclareMathOperator*{\argmin}{arg\,min}

\newcommand*\Let[2]{\State #1 $\gets$ #2}
\global\long\def\la{\langle}%
\global\long\def\ra{\rangle}%
\global\long\def\diag{\mathrm{diag}}%

\usepackage{acro}
\DeclareAcronym{MCMC}{short = MCMC , long = Markov chain Monte Carlo}
\DeclareAcronym{PDE}{short = PDE , long = partial differential equation}
\DeclareAcronym{CG}{short = CG , long = conjugate gradient}
\DeclareAcronym{SVD}{short = SVD , long = singular value decomposition}
\DeclareAcronym{EVD}{short = EVD , long = eigenvalue decomposition}
\DeclareAcronym{FITC}{short = FITC , long = fully independent training conditional}

\begin{document}

\title{Fast Approximate Solution of Stein Equations for Post-Processing of MCMC}
\author{Qingyang Liu$^1$, Heishiro Kanagawa$^1$, Matthew A. Fisher$^1$, Fran\c{c}ois-Xavier Briol$^2$, Chris. J. Oates$^{1,3}$ \\ 
\small $^1$Newcastle University, UK \\
\small $^2$University College London, UK \\
\small $^3$The Alan Turing Institute, UK }
\maketitle

\begin{abstract}
    Bayesian inference is conceptually elegant, but calculating posterior expectations can entail a heavy computational cost.
Monte Carlo methods
are reliable and supported by strong asymptotic guarantees, but do not leverage smoothness of the integrand.
Solving Stein equations has emerged as a possible alternative, providing a framework for numerical approximation of posterior expectations in which smoothness can be exploited.
However, existing numerical methods for Stein equations are associated with high computational cost due to the need to solve large linear systems.
This paper considers the combination of iterative linear solvers and preconditioning strategies to obtain fast approximate solutions of Stein equations.
\end{abstract}

\section{Introduction}

Conditional probability underpins statistical inference, yet the actual calculation of conditional probabilities remains a challenging computational task.
Our focus is on probability distributions arising in Bayesian statistics, where one starts with the joint distribution of parameters $\mathbf{X}$ and data $\mathbf{Y}$, and seeks to compute the distribution of parameters $\mathbf{X}$ conditional upon the realised dataset $\mathbf{Y} = \mathbf{y}$.
Eschewing measurability considerations, we assume that this conditional distribution admits a Lebesgue density $p(\cdot)$ on $\mathbb{R}^d$.
From Bayes' theorem, this conditional density is given by
\begin{align}
    p(\mathbf{x}) := 
    p_{\mathbf{X}|\mathbf{Y}}(\mathbf{x}|\mathbf{y}) = \frac{p_{\mathbf{X}}({x}) p_{\mathbf{Y}|\mathbf{X}}(\mathbf{y}|\mathbf{x})}{p_{\mathbf{Y}}(\mathbf{y})} \label{eq: Bayes}
\end{align}
where $p_{\mathbf{X}}(\cdot)$ denotes a marginal density of $\mathbf{X}$ (`the prior'), $p_{\mathbf{Y}|\mathbf{X}}(\cdot|\mathbf{x})$ denotes a probability mass function or density for $\mathbf{Y}$ conditional on $\mathbf{X} = \mathbf{x}$ (`the likelihood'), and $p_{\mathbf{Y}}(\cdot)$ denotes a marginal mass function or density for $\mathbf{Y}$ (`the marginal likelihood').
Practical applications of Bayesian statistics are characterised by explicit access to $p_{\mathbf{X}}(\cdot)$ and $p_{\mathbf{Y}|\mathbf{X}}(\cdot|\mathbf{x})$, but not $p_{\mathbf{Y}}(\cdot)$ or $p_{\mathbf{X}|\mathbf{Y}}(\cdot | \mathbf{y})$.
The conditional distribution $p_{\mathbf{X}|\mathbf{Y}}(\cdot|\mathbf{y})$ is termed the \emph{posterior} and is the object of scientific interest, from which conclusions are deduced.
Armed with the prior and the likelihood, in simple situations one can directly calculate the marginal likelihood $
    p_{\mathbf{Y}}(\mathbf{y}) = \int_{\mathbb{R}^d} p_{\mathbf{X}}(\mathbf{x}) p_{\mathbf{Y}|\mathbf{X}}(\mathbf{y}|\mathbf{x}) \; \mathrm{d}\mathbf{x}$
and thus obtain the posterior density $p_{\mathbf{X}|\mathbf{Y}}(\cdot|\mathbf{y})$ via \eqref{eq: Bayes}.
However, in all but the simplest situations $p_{\mathbf{Y}}(\mathbf{y})$ represents an analytically intractable integral which is often also numerically intractable, due to the highly localised nature of the integrand in the context of an informative likelihood.

Numerical approximation of posterior distributions has been a central research topic in statistics since the advent of \ac{MCMC}.
There are now myriad ingenious ways to obtain asymptotically exact approximations to the posterior \eqref{eq: Bayes}.
For example, \ac{MCMC} methods \cite{fearnhead2024scalable}, sequential Monte Carlo methods \cite{chopin2020introduction}, variational methods \cite{blei2017variational}, and gradient flow methods \cite{chen2023gradient}.
Among these, sampling-based methods are widely considered state-of-the-art \cite{bhattacharya2024grand}.
This is due in perhaps equal part to their widely understood (asymptotic) theoretical guarantees \cite{meyn2012markov} and the availability of production-level software \cite{carpenter2017stan}.
At a high-level, sampling-based methods output a sequence $(\mathbf{x}_n)_{n \in \mathbb{N}} \subset \mathbb{R}^d$ such that the distribution of $\mathbf{x}_n$ converges in an appropriate sense to the posterior \eqref{eq: Bayes}, and such that $\mathbf{x}_{m}$ is approximately independent from $\mathbf{x}_n$ whenever the indices $m$ and $n$ are sufficiently far apart.
Omitting technical details, one can make rigorous sense of these properties to guarantee the almost sure convergence of averages 
\begin{align}
\frac{1}{N} \sum_{n=1}^N f(\mathbf{x}_i) \rightarrow \int f(\mathbf{x}) p(\mathbf{x}) \mathrm{d}\mathbf{x} \label{eq: mcmc estimator}
\end{align} 
for all $f$ such that $ \int f^2(\mathbf{x}) p(\mathbf{x}) \mathrm{d}\mathbf{x} < \infty$; see e.g. Chapter 17 of \cite{meyn2012markov}.
Further, the approximation error is typically $O_P(N^{-1/2})$, whose exponent is dimension-independent. 
Considering that parameters of statistical models are often high-dimensional (i.e. $\mathbf{X} \in \mathbb{R}^d$, $d \gg 1$), the dimension-independent convergence rate is usually regarded as a positive attribute of sampling methods.
However, there are two scenarios where sampling methods are sub-optimal:
\begin{enumerate}
    \item the integrand $f$ is in some sense `smooth relative to the dimension $d$'
    \item evaluation of $f$ incurs a non-negligible cost.
\end{enumerate}
In the first case, by analogy with established cubature methods, one might expect to obtain a faster convergence rate by leveraging the smoothness of the integrand \cite{Novak2024}.
The challenge here is that cubature methods are not directly applicable in the Bayesian context, due to the implicit characterisation of the posterior distribution via \eqref{eq: Bayes}.
In the second case, the relatively slow error decay of sampling-based methods means that a potentially high cost is paid to obtain an accurate numerical approximation to the integral.
Yet this rate of convergence is characteristic to all Monte Carlo methods, and it is difficult to see how any meaningful acceleration can be achieved within the class of sampling methods, except for perhaps improving the rate constant.
A promising solution is to embed ideas from quasi Monte Carlo into \ac{MCMC} \cite{owen2005quasi}; however, this requires bespoke sampling algorithms to be used, and at present these do not not enjoy comparable software support to \ac{MCMC}.

Solving Stein equations has  emerged as a promising solution to both of the problems just discussed.
In brief, Stein equations cast the task of calculating a posterior expectation as the task of solving a \ac{PDE}.
Though solving a \ac{PDE} appears to be a more challenging computational task, a key observation is that \ac{PDE} solvers can directly exploit smoothness. 
This enables improved convergence rates for numerical approximations, which can in turn reduce the cost required to achieve a certain level of numerical precision in approximating posterior expected quantities of interest.

To date, several research groups have explored collocation and closely related numerical methods for solving Stein equations, where the sequence of collocation nodes are chosen to be the output $(\mathbf{x}_n)_{1 \leq n \leq N}$ from a sampling method, usually \ac{MCMC} \cite{oates2017control,oates2019convergence,barp2022riemann,si2020scalable,belomestny2020variance,south2022semi,south2022postprocessing,south2023regularized,sun2023vector,leluc2023speeding,belomestny2024theoretical}.
Despite theoretical support and encouraging empirical demonstrations, the computational complexity of these methods can often preclude their application to typically-sized \ac{MCMC} output.
For example, the computational complexity of kernel-based collocation methods is typically $O(N^3)$, while $N = 10^4$ or $N = 10^5$ samples are routinely produced using \ac{MCMC}.
There exist numerical approximation techniques for scaling collocation methods to large $N$, both in the \ac{PDE} literature and in the kernel methods literature, but a detailed and objective assessment of their suitability for solving Stein equations has yet to be performed. 

The aims of this article are threefold.
First, we aim for a self-contained exposition of collocation methods for solving Stein equations and their application to computing posterior expectations.
Second, we aim to review existing preconditioning techniques for scaling collocation-type methods to large datasets, with a critical discussion of their suitability for solving Stein equations in our motivating context.
Third, we aim to present an objective empirical assessment of the performance gains than can be achieved using the strategies which will be discussed.
This paper is accompanied by Online Appendices, which can be accessed at \url{https://arxiv.org/abs/2501.06634}.

\section{Background}

We now recall how Stein equations can be used to post-process \ac{MCMC} output.
\Cref{subsec: stein eq} introduces Stein equations, and
\Cref{subsec: using} discusses how approximate solutions of Stein equations enable approximate computation of posterior expectations.
Numerical methods, including collocation methods, are discussed in \Cref{subsec: solving stein}.
The main numerical techniques that we consider are iterative linear solvers and preconditioning, and we briefly recall how these can be applied for collocation in \Cref{subsec: precond}.

\subsection{Stein Equations}
\label{subsec: stein eq}

Stein equations provide a mathematical link between a posterior expected quantity of interest and the solution of a \ac{PDE}.
They were introduced in a special case by Charles Stein in \cite{stein1972bound}, where they were initially just a theoretical tool.
There are now a plethora of different approaches to construct Stein equations with both theoretical and computational purposes in mind \cite{anastasiou2023stein}, but we focus only on the \emph{canonical} Stein equation in this work.
Let $\nabla$ denote the gradient and $\nabla \cdot$ denote the divergence operators in $\mathbb{R}^d$.
Given a probability density function $p$ on $\mathbb{R}^d$, an integrand $f : \mathbb{R}^d \rightarrow \mathbb{R}$ of interest, and assuming that the gradient $\nabla \log p : \mathbb{R}^d \rightarrow \mathbb{R}^d$ is well-defined, the canonical \emph{Stein equation} (also known as the \emph{Langevin Stein equation}) is the first-order \ac{PDE}
\begin{align}
    f(\mathbf{x}) = c + \frac{1}{p(\mathbf{x})} (\nabla \cdot (p \mathbf{u}))(\mathbf{x}) , \qquad \mathbf{x} \in \mathbb{R}^d, \label{eq: stein}
\end{align}
whose solutions are defined as pairs $(c,\mathbf{u})$ where $c \in \mathbb{R}$ is a constant and $\mathbf{u} : \mathbb{R}^d \rightarrow \mathbb{R}^d$ is a differentiable vector field. 
The existence of solutions to the Stein equation is a deep and technical subject beyond this paper, but we note that if $\mathbf{u}$ is a gradient field $\mathbf{u} = \nabla v$ then existence of a solution to the Stein equation can be deduced from existence of a solution to the (weighted) Poisson equation on $\mathbb{R}^d$. 
See e.g. Proposition 2 of \cite{si2020scalable} for further detail, or see Theorem 1 of \cite{barp2022riemann} for the simpler case where the domain is a closed compact Riemannian manifold.

There are two key observations explaining the relevance of the Stein equation:

\begin{enumerate}
    \item \textit{Computability:}
From the product rule for differentiation, we can re-express the Stein equation as $f(\mathbf{x}) = c + (\nabla \cdot \mathbf{u})(\mathbf{x}) + \mathbf{u}(\mathbf{x}) \cdot (\nabla \log p)(\mathbf{x})$.
The term $\nabla \log p$ completely characterises the density $p$, and can be computed for posterior distributions directly from \eqref{eq: Bayes} without evaluation of intractable integrals, since
\begin{align}
    (\nabla \log p)(\mathbf{x}) = (\nabla \log p_{\mathbf{X}})(\mathbf{x}) + (\nabla \log L)(\mathbf{x}) \label{eq: score}
\end{align}
where $L(\cdot) = p_{\mathbf{Y}|\mathbf{X}}(\mathbf{y}|\cdot)$ is the likelihood. 
Indeed, most modern \ac{MCMC} algorithms exploit \eqref{eq: score} as part of their Markov transition kernel (for example, in Metropolis--Hastings one can propose a candidate state for the Markov chain by moving in a direction of increasing log-probability gradient), and as such the gradients $(\nabla \log p)(\mathbf{x}_n)$  can be cached along the
sample path, so that they are available for use in post-processing at no additional computational cost.

\item \textit{Integral Approximation:}
If $(c,\mathbf{u})$ is a solution to the Stein equation, then from an appropriately general formulation of the divergence theorem
\begin{align}
    \int f(\mathbf{x}) p(\mathbf{x}) \; \mathrm{d}\mathbf{x} = c +  \int (\nabla \cdot (p\mathbf{u}))(\mathbf{x}) \; \mathrm{d}\mathbf{x} =c+0 =c \label{eq: integrate to 0}
\end{align}
provided $p\mathbf{u}$ and $\nabla \cdot (p\mathbf{u})$ are both integrable on $\mathbb{R}^d$; see \cite{barp2022targeted} or Proposition 3 of \cite{oates2022minimum}.
This indicates that $c$ is precisely the value of the posterior expectation of interest, suggesting that if we can numerically approximate a solution $(c,\mathbf{u})$ of the Stein equation then we can read off an approximation to the corresponding posterior expectation as well.
Further details are provided next.

\end{enumerate}

\subsection{Using Approximate Solutions of Stein Equations}
\label{subsec: using}

Armed with an approximate solution $(\tilde{c} , \tilde{\mathbf{u}})$ to the canonical Stein equation \eqref{eq: stein}, there are two main ways in which the approximation can be leveraged to approximate the posterior integral $\int f(\mathbf{x}) p(\mathbf{x}) \; \mathrm{d}\mathbf{x}$ of interest.

\begin{enumerate}
    \item \textit{Point Estimator:}
First, one may directly take $\tilde{c}$ as an approximation to the integral of interest, motivated by \eqref{eq: integrate to 0}.
This approach was considered in works such as \cite{barp2022riemann,south2022semi,Briol2017SMCKQ,Chen2024}, where it was shown to be effective providing the approximating class is rich enough that the ground truth $(c , \mathbf{u})$ can be (in an appropriate sense) consistently approximated.

\item \textit{Control Variate:}
One can use the corresponding approximation to $f$ as a control variate to construct a reduced-variance alternative to the estimator in \eqref{eq: mcmc estimator}
$$
\frac{1}{N} \sum_{n=1}^N \left[ f(\mathbf{x}_n) - \beta \tilde{f}(\mathbf{x}_n) \right] + \beta \underbrace{ \int \tilde{f}(\mathbf{x}) p(\mathbf{x}) \; \mathrm{d}\mathbf{x} }_{= \tilde{c}} , \qquad \tilde{f} = \tilde{c} + \frac{1}{p} \nabla \cdot(p \tilde{\mathbf{u}})
$$
for some $\beta \in \mathbb{R}$.
This was the approach considered by authors such as \cite{belomestny2020variance,belomestny2024theoretical}.
The variance of this estimator with respect to the randomness of sampling $(\mathbf{x}_n)_{1 \leq n \leq N}$ may be smaller than that of the usual \ac{MCMC} estimator. 
However, the analysis of this strategy is complicated by the reality that $\tilde{f}$ would typically be constructed using the same samples $(\mathbf{x}_n)_{1 \leq n \leq N}$, and these samples are typically correlated samples since they arise from \ac{MCMC}.

\end{enumerate}

\noindent For simplicity, the point estimator approach will be considered in the present manuscript, but we note that fundamental task of numerically solving the Stein equation is common to both approaches.

\subsection{Solving the Canonical Stein Equation}
\label{subsec: solving stein}

Several strategies for numerical solution of the Stein equation can be conceived.
A reflexive strategy is to assume a gradient field $\mathbf{u} = \nabla v$ and then use existing numerical methods designed for second-order elliptic \acp{PDE}.
Among such methods, Galerkin methods are popular and well-understood, with finite element bases often a practical gold-standard. 
However, there are features of the Stein equation for which Galerkin methods are not well-suited.
First, the equation is often defined on an unbounded domain, which is somewhat non-standard. 
Although it may be practically sufficient to solve the Stein equation in regions where most of the probability mass of $p$ is contained, this region is typically unknown at the outset.
Second, the dimension of the domain can often be large (i.e. $d \gg 1$) so that the number of basis elements needed to appropriately resolve the solution can be prohibitively large in general.
Third, the integration over basis elements that is required to obtain a stiffness matrix involves a weighted integral with respect to the density $p$, up to a normalisation constant, introducing a circularity since integration with respect to $p$ is the motivating task.

As a result, several research groups have instead explored collocation (or \emph{meshless}) methods for numerical solution of the Stein equation.
Collocation methods are characterised as seeking a strong solution to the \ac{PDE}, requiring only that the quantities appearing in the \ac{PDE} can be pointwise evaluated, and scaling favourably with dimension by circumventing the explicit construction of a mesh \cite{fasshauer2008meshfree,chen2017meshfree}. Among collocation methods for \acp{PDE}, our focus is on the popular \emph{symmetric collocation} method \cite{fasshauer1996solving}, since this has elegant connections with minimum norm interpolation and kernel methods that we wish to exploit. To introduce these, we first consider the simplified setting of a linear \ac{PDE} of the form $f = \mathcal{L} v$, where the strong solution $v : \mathbb{R}^d \rightarrow \mathbb{R}$ is assumed to uniquely exist as an element of $\mathcal{H}(k)$, the Hilbert space reproduced by a symmetric and positive definite kernel $k : \mathbb{R}^d \times \mathbb{R}^d \rightarrow \mathbb{R}$. 
Assuming the linear functionals $v \mapsto (\mathcal{L}v)(\mathbf{x}_n)$ are continuous, the symmetric collocation method approximates $v$ as
\begin{align}
    v_N(\mathbf{x}) = \sum_{n=1}^N w_n \mathcal{L}_2 k( \mathbf{x} , \mathbf{x}_n ) \label{eq: representer}
\end{align}
where $(\mathbf{x}_n)_{1 \leq n \leq N}$ are the \emph{collocation nodes}, and $\mathcal{L}_i$ denotes the action of the operator on the $i$th argument of a bivariate functional; see Chapter 16 in \cite{Wendland2004}.
The weights $\mathbf{w}$ are obtained from solving the linear system of equations $\mathbf{A} \mathbf{w} = \mathbf{b}$ where
\begin{align*}
    \mathbf{A} = [A_{i,j}]_{1 \leq i,j \leq N}, \qquad A_{i,j} = (\mathcal{L}_1 \mathcal{L}_2 k)(\mathbf{x}_i,\mathbf{x}_j),
\end{align*}
is called the \emph{collocation matrix} and
\begin{align*}
    \mathbf{b} = [b_i]_{1 \leq i \leq N}, \qquad b_i = f(\mathbf{x}_i) .
\end{align*}
The symmetric collocation method is rather natural and arises from several different perspectives.
One perspective is minimal norm interpolation, where \eqref{eq: representer} is the solution to the variational problem
\begin{align}
    \argmin_{v \in \mathcal{H}(k)} \|v\|_{\mathcal{H}(k)} \; \text{such that} \; f(\mathbf{x}_n) = (\mathcal{L}v)(\mathbf{x}_n) , \; n = 1,\dots,N, \label{eq: collo as vari}
\end{align}
see Theorem 16.1 of \cite{Wendland2004}. 
Though the Stein equation \eqref{eq: stein} does not exactly fit into this set-up, due to the presence of the unknown constant $c$ and the fact that the function $\mathbf{u}$ appearing in the Stein equation is a vector field, we explain in \Cref{sec: methods} how the numerical solution of the Stein equation \eqref{eq: stein} can also be couched as a variational problem in a similar spirit to \eqref{eq: collo as vari} with the \emph{Langevin Stein operator} $\mathcal{L}\mathbf{u} = \frac{1}{p} \nabla \cdot (p \mathbf{u})$, and thus approximately solved using a collocation-type method.

\subsection{Conjugate Gradients and Preconditioning of Collocation Methods}
\label{subsec: precond}

Under appropriate regularity conditions, the matrix $\mathbf{A} \in \mathbb{R}^{N \times N}$ appearing in collocation methods has a well-defined inverse $\mathbf{A}^{-1}$; see Chapter 16 in \cite{Wendland2004}.
However, the matrix $\mathbf{A}$ can become severely ill-conditioned for moderate-to-large values of $N$, meaning that $\mathbf{A}^{-1}$ cannot be accurately computed using a direct method.
Further, the memory requirement of simply storing $\mathbf{A}$ in typically-sized RAM on a personal computer can be prohibitive when $N$ is larger than a few thousand. 
As such, iterative linear solvers that require only the \emph{action} of $\mathbf{A}$ in the form of a matrix-vector product, together with preconditioning strategies, are routinely employed in combination with the symmetric collocation method. 
Let $\lambda_{\text{min}}(\mathbf{A})$ and $ \lambda_{\text{max}}(\mathbf{A})$ denote the minimum and maximum eigenvalues of $\mathbf{A}$, and  $\mathrm{cond}(\mathbf{A})= \lambda_{\text{max}}(\mathbf{A})\lambda_{\text{min}}(\mathbf{A})^{-1}$ denote the condition number of $\mathbf{A}$, with large values representing poor conditioning. 
Practically, one first seeks a symmetric and cheaply invertible \emph{preconditioner} matrix of the form $\mathbf{M} = \mathbf{E} \mathbf{E}^\top$ for some $\mathbf{E} \in \mathbb{R}^{N \times N}$ such that $\mathrm{cond}(\mathbf{E}^{-1} \mathbf{A} \mathbf{E}^{-\top}) \ll \mathrm{cond}(\mathbf{A})$, and then one approximately solves 
\begin{align}
\tilde{\mathbf{A}} \tilde{\mathbf{w}} = \tilde{\mathbf{b}} , \qquad \tilde{\mathbf{A}} = \mathbf{E}^{-1} \mathbf{A} \mathbf{E}^{-\top} , \qquad \tilde{\mathbf{b}} = \mathbf{E}^{-1} \mathbf{b} \label{eq: pre lin sys}
\end{align}
using $m$ iterations of an iterative linear solver, such as the \ac{CG} method \cite{hestenes1952methods}, returning an approximation $\tilde{\mathbf{w}}_m$ to $\tilde{\mathbf{w}}$. 
The preconditioned linear system $\tilde{\mathbf{A}} \tilde{\mathbf{w}} = \tilde{\mathbf{b}}$ is equivalent to the original linear system $\mathbf{A} \mathbf{w} = \mathbf{b}$, since one can extract $\mathbf{w}_m = \mathbf{E}^{-\top} \tilde{\mathbf{w}}_m$ as a numerical approximation to the solution vector $\mathbf{w}$ of interest.
Further, since the preconditioned matrix $\tilde{\mathbf{A}}$ has a smaller condition number compared to the original matrix $\mathbf{A}$, fewer iterations of an iterative method will typically be required \cite{nocedal1999numerical}. 
In particular, Theorem 11.3.3 of \cite{Golub2013} shows that for a sequence of approximations $\tilde{\mathbf{w}}_1, \tilde{\mathbf{w}}_2,\ldots$ produced using the \ac{CG} method applied to \eqref{eq: pre lin sys},
\begin{align}
    \|\tilde{\mathbf{w}}_m-\tilde{\mathbf{w}}\|_{\tilde{\mathbf{A}}} \leq \left(1-\frac{1}{\text{cond}(\tilde{\mathbf{A}})}\right)^{\frac{1}{2}} \|\tilde{\mathbf{w}}_{m-1} - \tilde{\mathbf{w}} \|_{\tilde{\mathbf{A}}} , \qquad \|\mathbf{z}\|_{\tilde{\mathbf{A}}} = (\mathbf{z}^\top \tilde{\mathbf{A}} \mathbf{z})^{1/2} , 
\end{align}
suggesting that a smaller condition number for $\tilde{\mathbf{A}}$ entails faster convergence of the \ac{CG} method. 
The ideal preconditioner is one for which $\mathrm{cond}(\tilde{\mathbf{A}})=1$. This can in theory be achieved by taking $\mathbf{E} = \mathbf{A}^{1/2}$, but this is as hard as solving the numerical system itself. Instead, it is common to use approximations of this quantity which are easier to invert.
For example, Sec. 3.1.2 of \cite{fasshauer1999solving} advocated a Jacobi preconditioner, meaning  $\mathbf{E} = \mathrm{diag}(\mathbf{A})^{1/2}$, presenting empirical evidence that even this simple preconditioning strategy can markedly improve the overall condition of the system of linear equations that must be solved.
Likewise, the importance of preconditioning for large-scale kernel methods is well-understood \cite{cutajar2016preconditioning}, with Nystr\"{o}m-based preconditioning \cite{rudi2017falkon,abedsoltan2024nystrom} being popular in the field.
Yet, a comparison of the suitability and effectiveness of different preconditioning techniques for numerical solution of the Stein equation has yet to be performed.
This is our focus next.

\section{Fast Approximate Solutions of Stein Equations}
\label{sec: methods}

The numerical solution of the Stein equation \eqref{eq: stein} is described in a variational form in \Cref{subsec: variational}. Subject to practical considerations discussed in \Cref{subsec: practical}, this leads to a linear system of equations which can then be further preconditioned.
A range of different preconditioning techniques are described in \Cref{subsec: preconditioners}.

\subsection{A Variational Formulation}
\label{subsec: variational}

This section explains how the basic form of the symmetric collocation method from \Cref{subsec: solving stein} can be extended to the case of the Stein equation \eqref{eq: stein}, following the approach of \cite{oates2017control,oates2019convergence,barp2022riemann}.
Again we let $k : \mathbb{R}^d \times \mathbb{R}^d \rightarrow \mathbb{R}$ be a symmetric and positive definite kernel, which we will call the \emph{base kernel}, and now we let $\mathcal{U} = \mathcal{H}(k) \times \cdots \times \mathcal{H}(k)$ denote the Cartesian product of $d$ copies of $\mathcal{H}(k)$, with norm satisfying $\|\mathbf{u}\|_{\mathcal{U}}^2 = \sum_{i=1}^d \|u_i\|_{\mathcal{H}(k)}^2$.
Taking inspiration from collocation methods for \acp{PDE}, we cast approximate numerical solution of the Stein equation \eqref{eq: stein} as the variational problem
\begin{align}
\argmin_{c \in \mathbb{R} , \mathbf{u} \in \mathcal{U}} \|\mathbf{u}\|_{\mathcal{U}} \; \text{such that} \; f(\mathbf{x}_n) = c + \frac{1}{p(\mathbf{x}_n)} (\nabla \cdot (p \mathbf{u}))(\mathbf{x}_n) , \qquad n = 1,\dots , N , \label{eq: Stein vari}
\end{align}
whose solution takes the form of a collocation-type method.
Indeed, $\mathbf{u} \in \mathcal{U}$ implies that $\frac{1}{p} \nabla \cdot (p\mathbf{u}) \in \mathcal{H}(k_p)$ where $k_p : \mathbb{R}^d \times \mathbb{R}^d \rightarrow \mathbb{R}$ is the \emph{Stein reproducing kernel}
\begin{align}
    k_p(\mathbf{x},\mathbf{x}') & = (\nabla_1 \cdot \nabla_2 k)(\mathbf{x},\mathbf{x}') + (\nabla_1 k)(\mathbf{x},\mathbf{x}') \cdot (\nabla \log p)(\mathbf{x}') \label{eq: stein kernel} \\
    & \quad + (\nabla \log p)(\mathbf{x}) \cdot (\nabla_2 k)(\mathbf{x},\mathbf{x}') + k(\mathbf{x},\mathbf{x}') (\nabla \log p)(\mathbf{x}) \cdot (\nabla \log p)(\mathbf{x}') \nonumber
\end{align}
and $\|v\|_{\mathcal{H}(k_p)} = \inf\{ \|\mathbf{u}\|_{\mathcal{U}} : v =  \frac{1}{p} \nabla \cdot (p\mathbf{u}) , \; \mathbf{u} \in \mathcal{U} \}$; see Theorem 1 of \cite{oates2017control}.
Here we have used $\nabla_i$ to denote gradient and $\nabla_i \cdot$ to denote the divergence with respect to the $i$th argument.
Thus \eqref{eq: Stein vari} can be expressed as
\begin{align}
\argmin_{c \in \mathbb{R} , v \in \mathcal{H}(k_p)} \|v\|_{\mathcal{H}(k_p)} \; \text{such that} \; f(\mathbf{x}_n) = c + v(\mathbf{x}_n) , \qquad n = 1,\dots , N . \label{eq: Stein vari 2}
\end{align}
The solution $(c_N,v_N)$ to the variational problem \eqref{eq: Stein vari 2} has an explicit closed form, which is summarised in the following result:

\begin{proposition}
\label{prop: explicit form}
Assume that the Stein kernel $k_p$ in \eqref{eq: stein kernel} is well-defined, and let the collocation nodes $(\mathbf{x}_n)_{1 \leq n \leq N}$ be distinct.
Then \eqref{eq: Stein vari 2} has a unique solution $(c_N,v_N)$ with the constant $c_N$ admitting the explicit closed form
\begin{align}
c_N = \frac{\mathbf{f}^\top \mathbf{K}_p^{-1} \mathbf{1}}{\mathbf{1}^\top \mathbf{K}_p^{-1} \mathbf{1}} , \qquad \mathbf{f} = [f(\mathbf{x}_i)]_{1 \leq i \leq N}, \qquad \mathbf{K}_p = [k_p(\mathbf{x}_i , \mathbf{x}_j) ]_{1 \leq i , j \leq N} ,   \label{eq: cN}
\end{align}
where $\mathbf{1} = [1,\dots,1]^\top \in \mathbb{R}^N$.
\end{proposition}

\noindent As this result is central to the paper, we provide a self-contained proof in the Online Appendix.
Several comments are immediately in order:
\begin{enumerate}
\item \Cref{prop: explicit form} assumes the collocation nodes are distinct, so that $\mathbf{K}_p$ can be inverted.
If the collocation nodes arise as output from Metropolis--Hastings \ac{MCMC}, then we will need to manually discard duplicate samples to proceed.
One might worry that the thinned samples are no longer $p$-invariant. 
However, $p$-invariance of the \ac{MCMC} output is not a necessary condition for consistency of the estimator $c_N$; essentially, all information about $p$ is contained in the Stein reproducing kernel; see e.g. Appendix L of \cite{south2022semi}.
Alternative sampling methods, which may generate collocation nodes better-suited to numerical solution of the Stein equation, are discussed in \cite{wang2024stein}, but to limit scope here we consider only collocation nodes that are generated using $p$-invariant \ac{MCMC}.

\item The resulting linear system that we need to solve is 
\begin{equation}
\mathbf{K}_p \mathbf{w} = \mathbf{1} . \label{eq: linear-system}
\end{equation}
Due to the number $N$ of samples that define this linear system, instantiating the matrix $\mathbf{K}_p$ in RAM can incur a prohibitive $O(N^2)$ computational cost.
As such, iterative methods that use only the action of $\mathbf{K}_p$ on vectors are strongly preferred.
To emphasise this requirement, we let $\mathrm{Act}_{\mathbf{K}_p} : \mathbb{R}^N \rightarrow \mathbb{R}^N$ denote the map $\mathbf{v} \mapsto \mathbf{K}_p \mathbf{v}$.
The computational complexity of evaluating $\mathrm{Act}_{\mathbf{K}_p}$ is $O(N^2)$, but the storage complexity is now $O(N)$.
Pseudocode for memory-efficient computation of $\mathrm{Act}_{\mathbf{K}_p}$ is presented in the Online Appendix.

\item The performance of iterative linear solvers such as the \ac{CG} method depends strongly on the condition number of the matrix $\mathbf{K}_p$.
Depending on the base kernel $k$ appearing in \eqref{eq: stein kernel}, the eigenvalue decay of $\mathbf{K}_p$ can be rapid, leading to $\mathbf{K}_p$ being severely ill-conditioned.

\end{enumerate}

\noindent Together these considerations  motivate the use of preconditioned iterative methods for \eqref{eq: linear-system} in which only the action of $\mathbf{K}_p$ is required.
For the experiments that we report in \Cref{sec: empirical} we use \ac{CG} as the iterative linear solver, as this is most widely-used.
Different preconditioning strategies for \eqref{eq: linear-system} are discussed in \Cref{subsec: preconditioners}.

\subsection{Practical Considerations}
\label{subsec: practical}

This section summarises the main practical considerations associated with numerical solution of the Stein equation; the choice of kernel, monitoring performance, and memory-efficient numerical preconditioning.

\subsubsection{Choice of Kernel}
\label{subsubsec: kernel choice}

As with all kernel methods, it is important to select a kernel that is appropriate for the task at hand.
In our case, this amounts to selecting an appropriate base kernel $k$ in \eqref{eq: stein kernel}.
For example, we might consider a base kernel of the form 
\begin{align}
k(\mathbf{x},\mathbf{x}') = \varphi\left( \frac{\mathbf{x} - \mathbf{x}'}{\ell} \right) \label{eq: trans invar}
\end{align}
where $\ell > 0$ is a length scale to be specified.
Standard techniques, such as cross-validation, provide a useful rationale for how $\ell$ can be selected.
However, these techniques require solving the Stein equation on subsets of the collocation nodes, which gives rise again to the motivating computational task.
In particular, larger length scales $\ell$ tend to result in matrices $\mathbf{K}_p$ that are more ill-conditioned.
Thus any computational advantage of the preconditioning techniques that we will consider in \Cref{subsec: preconditioners} can also be brought to bear on the kernel choice task.

\subsubsection{Monitoring Performance}
\label{subsubsec: performance}

Suppose that $\mathbf{w}_m$ is an approximate solution of \eqref{eq: linear-system}, for example as obtained using $m$ iterations of the \ac{CG} method.
Then we may consider the corresponding approximation $c_{N,m} = (\mathbf{f}^\top \mathbf{w}_m) / (\mathbf{1}^\top \mathbf{w}_m)$ to $c_N$ in \eqref{eq: cN}.
Let $f = c + v$ with $c \in \mathbb{R}$ and $v \in \mathcal{H}(k_p)$.
Noting that $c_{N,m}$ is a normalised cubature rule, an application of the reproducing property and Cauchy--Schwarz shows that the error of the estimator $c_{N,m} \equiv c_{N,m}(f)$ can be explicitly bounded:
\begin{align*}
     \left| c_{N,m}(f) - \int f(\mathbf{x}) p(\mathbf{x}) \; \mathrm{d}\mathbf{x} \right|  = \left| c_{N,m}(v) - \int v(\mathbf{x}) p(\mathbf{x}) \; \mathrm{d}\mathbf{x} \right| \nonumber \leq  \|v\|_{\mathcal{H}(k_p)} \sigma(\mathbf{w}_m) 
\end{align*}
where 
\begin{align}
\sigma(\mathbf{w}_m) 
    = \sup_{\|v\|_{\mathcal{H}(k_p)} \leq 1 } \left| c_{N,m}(v) - \int v(\mathbf{x}) p(\mathbf{x}) \; \mathrm{d}\mathbf{x} \right| 
    = \frac{ (\mathbf{w}_m^\top \mathbf{K}_p \mathbf{w}_m )^{1/2} }{\mathbf{1}^\top \mathbf{w}_m } .   \label{eq: performance bound}
\end{align}
Although the term $\|v\|_{\mathcal{H}(k_p)}$ will be unknown in general, one can monitor the worst-case error $\sigma(\mathbf{w}_m)$, providing a proxy for estimator performance that can be explicitly computed in $O(N^2)$ time and $O(N)$ storage cost.

\subsubsection{Memory-Efficient Preconditioning the Canonical Stein Equation}
\label{subsubsec: precondition}

Motivated by the discussion of preconditioning collocation methods in \Cref{subsec: precond}, here we consider analogous preconditioning for the Stein equation.
That is, we first solve
$$
\tilde{\mathbf{K}}_p \tilde{\mathbf{w}} = \mathbf{E}^{-1} \mathbf{1} , \qquad \tilde{\mathbf{K}}_p = \mathbf{E}^{-1} \mathbf{K}_p \mathbf{E}^{-\top} 
$$
for $\tilde{\mathbf{w}}$ using an iterative method, and then solve $\mathbf{w} = \mathbf{E}^{-\top} \tilde{\mathbf{w}}$.
The preconditioner is $\mathbf{M} = \mathbf{E} \mathbf{E}^\top$, and the hope is that $\tilde{\mathbf{K}}_p$ is better conditioned than $\mathbf{K}_p$.
Pseudocode for preconditioned \ac{CG} applied to the Stein equation is presented in the Online Appendix.
A notable feature of preconditioned \ac{CG} is that it requires only the action of $\mathbf{M}^{-1}$, denoted $\mathrm{Act}_{\mathbf{M}^{-1}} : \mathbb{R}^N \rightarrow \mathbb{R}^N$, rather than the matrix $\mathbf{M}$ itself, which permits memory efficient computation in a similar manner to that previously discussed.
In the next section we explore various different choices for the preconditioner matrix $\mathbf{M}$.

\subsection{Preconditioners}
\label{subsec: preconditioners}

The literature on collocation and kernel methods contains different approaches to preconditioning, and a non-exhaustive selection of representative examples will be discussed.

\subsubsection{Jacobi Preconditioning}

Following the suggestion of \cite{fasshauer1999solving}, we could consider Jacobi preconditioning with $\mathbf{E} = \mathrm{diag}(\mathbf{K}_p)^{1/2}$. 
Let $\Delta = \nabla \cdot \nabla$ denote the Laplacian on $\mathbb{R}^d$.
For base kernels of the form \eqref{eq: trans invar}, from \eqref{eq: stein kernel} we have
\begin{align}
[\mathbf{K}_p]_{i,i} = k_p(\mathbf{x}_i,\mathbf{x}_i) = - \frac{ (\Delta \varphi)(\mathbf{0}) }{ \ell^2 } + \varphi(\mathbf{0}) \| (\nabla \log p)(\mathbf{x}_i) \|^2 \label{eq: diagonal}
\end{align}
so that $[\mathbf{E}]_{i,i} \asymp \| (\nabla \log p)(\mathbf{x}_i) \|$ in the tail.
For example, for a Gaussian posterior with mean $\bm{\mu}$ and covariance $\bm{\Sigma}$, we would have $[\mathbf{E}]_{i,i} \asymp \| \bm{\Sigma}^{-1/2} (\mathbf{x}_i - \bm{\mu}) \|$, meaning that Jacobi preconditioning is mitigating the negative effect of states $\mathbf{x}_i$ with low posterior probability on the conditioning of the matrix $\mathbf{K}_p$. 
On the other hand, for $\ell \rightarrow 0$ the first term in \eqref{eq: diagonal} dominates and $\mathbf{E}$ converges to a multiple of the identity matrix, meaning that no meaningful preconditioning is achieved.
An extension of Jacobi preconditioning is \emph{block} Jacobi, where $\mathbf{E}$ is taken to be a block diagonal matrix whose blocks are of size $b \times b$, for some $b$ to be specified, and coincide with the corresponding elements of $\mathbf{K}_p$. 

\subsubsection{Nystr\"{o}m Preconditioning}\label{subsec:nyst-precond}

The Nystr\"{o}m method \cite{nystrom1930praktische}
samples a subset of $n \leq N$ collocation nodes (specified by indices $1\leq s_1 < \dots < s_n \leq N$, called the \emph{inducing points}) and uses these to approximate the full matrix $\mathbf{K}_p$.
To avoid cumbersome notation, we will also use the shorthand $\mathbf{K}$ for $\mathbf{K}_p$, where the $p$ subscript is omitted.
Specifically, the Nystr\"{o}m method approximates $\mathbf{K}$ by
\begin{align}
    \tilde{\mathbf{K}} = \mathbf{K}_{N,n}\mathbf{K}_{n,n}^{-1}\mathbf{K}_{n,N}, \qquad 
    \begin{array}{l} 
    \mathbf{K}_{n,n} = [k_p(\mathbf{x}_{s_i} , \mathbf{x}_{s_j})]_{1 \leq i , j \leq n} , \\
    \mathbf{K}_{N,n} = \mathbf{K}_{n,N}^\top = [k_p(\mathbf{x}_i , \mathbf{x}_{s_j})]_{1 \leq i \leq N , \; 1 \leq j \leq n} 
    \end{array} ,  \label{eq: first nystrom}
\end{align}
which can be interpreted as computing the similarity between inputs $\mathbf{x}_i$ and $\mathbf{x}_j$ only through the projection of their feature vectors onto the span of the feature vectors corresponding to inducing points.
To convert this approximation into a preconditioner, the Woodbury method is commonly used:
\begin{align}
\mathbf{M}^{-1} = (\mathbf{K}_{N,n}\mathbf{K}_{n,n}^{-1}\mathbf{K}_{n,N} + \eta \mathbf{I})^{-1}
= \eta^{-1}[\mathbf{I} - \mathbf{K}_{N,n}(\eta \mathbf{K}_{n,n}+\mathbf{K}_{n,N}\mathbf{K}_{N,n})^{-1}\mathbf{K}_{n,N}]
\label{eq:woodbury_Nys}
\end{align}
where $\eta > 0$ is a parameter of the preconditioner (sometimes called a \emph{nugget}) to be specified.
The time complexity of calculating $\mathbf{M}^{-1}$ is $O(n^3 + n^2 N)$, significantly smaller than $O(N^3)$ for taking the inverse of $\mathbf{K}$ when $n \ll N$.
Nystr\"{o}m approximations have recently been considered in combination with Stein reproducing kernels in \cite{kalinke2024nystr}, though there the task of numerically solving the Stein equation was not considered.

\subsubsection{Nystr\"om Preconditioning and Diagonal Sampling}
The performance of Nystr\"{o}m-based methods can be sensitive to which $n$ inducing points are selected, and several methods for node selection have been developed. 
These methods can be classified into \emph{fixed sampling} and \emph{adaptive sampling}. 
Fixed sampling methods sample without replacement from a fixed probability distribution to select $n$ nodes, with the uniform distribution giving rise to the classic Nystr\"{o}m method. 
Adaptive sampling methods instead update the sampling scheme to ensure good coverage of the domain, typically by down-weighting nodes that are close to nodes which have already been selected. 
A trade-off between efficiency and accuracy exists for these sampling methods, as complex sampling methods spend more time selecting higher quality nodes for Nystr\"{o}m method.

For this manuscript, we focus on a fixed sampling method called \emph{diagonal sampling}. Diagonal sampling \cite{Drineas20052153} samples collocation nodes with probabilities proportional to $[\mathbf{K}_p]_{i,i}$ (note that the referenced work contains an erroneous square on the sampling probability, which was rectified in later work).
From \eqref{eq: diagonal}, we see that diagonal sampling preferentially samples nodes from the tail of $p$; i.e. samples are over-dispersed.
This idea is similar to that of \cite{wang2024stein}, who suggested obtaining the original node set $(\mathbf{x}_i)_{1 \leq i \leq N}$ by sampling from $q(\mathbf{x}) \propto \sqrt{k_p(\mathbf{x},\mathbf{x})} p(\mathbf{x})$ using \ac{MCMC}, which also leads to samples that are typically over-dispersed.

\subsubsection{Fully Independent Training Conditional}
The \ac{FITC} approach was introduced in \cite{quinonero2005unifying} as an approximation technique for computing with Gaussian processes, where $\mathbf{K}_p$ is interpreted as a covariance matrix.
To avoid cumbersome notation, we again use the shorthand $\mathbf{K}$ for $\mathbf{K}_p$, where the $p$ subscript is omitted.
The \ac{FITC} approach can be viewed as the Nystr\"{o}m approximation \eqref{eq: first nystrom} together with an additional correction term:
\begin{align*}
    \tilde{\mathbf{K}} = \mathbf{K}_{N,n}\mathbf{K}_{n,n}^{-1}\mathbf{K}_{n,N} + \diag(\mathbf{K} - \mathbf{K}_{N,n}\mathbf{K}_{n,n}^{-1}\mathbf{K}_{n,N})
\end{align*}
Here the second term is a diagonal matrix which serves as a correction term for the marginal variance that is not fully captured by the inducing points (i.e. the diagonal of $\tilde{\mathbf{K}}$ now coindices with the diagonal of $\mathbf{K}$). 
Denote $\mathbf{D} := \diag(\mathbf{K} - \mathbf{K}_{N,n}\mathbf{K}_{n,n}^{-1}\mathbf{K}_{n,N}) + \eta\mathbf{I}$. Since $\mathbf{D}$ is diagonal, we use Woodbury matrix inversion formula and obtain 
\begin{align*}
    \mathbf{M}^{-1} & = [ \mathbf{K}_{N,n}\mathbf{K}_{n,n}^{-1}\mathbf{K}_{n,N} + \underbrace{ \diag(\mathbf{K} - \mathbf{K}_{N,n}\mathbf{K}_{n,n}^{-1}\mathbf{K}_{n,N}) + \eta \mathbf{I} }_{=: \mathbf{D}} ]^{-1} \\
    & = \mathbf{D}^{-1} - \mathbf{D}^{-1}\mathbf{K}_{N,n}(\mathbf{K}_{n,n}+\mathbf{K}_{n,N}\mathbf{D}^{-1}\mathbf{K}_{N,n})^{-1}\mathbf{K}_{n,N}\mathbf{D}^{-1}
\end{align*}
at $O(n^3 + n^2N)$ computational cost.

\subsubsection{Randomised Nystr\"om Preconditioning} 
The randomised Nystr\"{o}m preconditioner \cite{frangella2023randomized} leverages the same idea as Nystr\"{o}m, but instead of sampling a small subset of rows and columns of the kernel matrix, it uses random projections to achieve a low-rank approximation. 
Let $\mathbf{\Omega} \in \mathbb{R}^{N\times n}$ have entries drawn independently from the standard Gaussian distribution. 
The kernel matrix is then approximated by 
$$\tilde{\mathbf{K}}_p \approx \mathbf{K}_p \mathbf{\Omega}(\mathbf{\Omega}^\top\mathbf{K}_p \mathbf{\Omega}+\eta \mathbf{I})^{-1}(\mathbf{K}_p\mathbf{\Omega})^\top.$$
An application of the Woodbury identity leads to the preconditioner
\begin{align}
\mathbf{M}^{-1} & = (\mathbf{K}_p\mathbf{\Omega}(\mathbf{\Omega}^\top\mathbf{K}_p\mathbf{\Omega} + \eta \mathbf{I} )^{-1}(\mathbf{K}_p\mathbf{\Omega})^\top)^{-1} \nonumber \\
& = \eta^{-1}[\mathbf{I} - \mathbf{K}_p\mathbf{\Omega}(\eta (\mathbf{\Omega}^\top\mathbf{K}_p\mathbf{\Omega})+(\mathbf{K}_p\mathbf{\Omega})^\top\mathbf{K}_p\mathbf{\Omega})^{-1}(\mathbf{K}_p\mathbf{\Omega})^\top].
\end{align}
The randomised Nystr\"{o}m preconditioner has time complexity $O(n^3 + n^2N)$, identical to the classic Nystr\"{o}m preconditioner, but this perspective naturally accommodates several generalisations beyond the Gaussian matrix $\bm{\Omega}$ that we have presented.
For example, the use of sparse Johnson-Lindenstrauss transforms has been considered \cite{demmel2023improved}. 
For the present purposes only the Gaussian case is considered.

\subsubsection{Randomised Nystr\"{o}m Eigenvalue Decomposition}
\label{sec:rand_nystrom}

The two–stage framework of \cite{halko2011finding} can be used to obtain a rank–$n$ approximation of the kernel matrix $\mathbf{K}_p$:
The first stage (\emph{range finding}) generates a matrix $\mathbf{\Omega} \in \mathbb{R}^{N \times n}$ whose entries are independently sampled from the standard Gaussian distribution, where $n$ is the target rank, then forms
$$ \mathbf Y = \bigl(\mathbf{K_p} \mathbf{K_p}^{\top}\bigr)^{q}\,
\mathbf{K_p}\,\mathbf{\Omega}, $$
where $q\geq 0$ power iterations enlarge the spectral gap\footnote{Indeed, given the eigendecomposition $\mathbf{K}_p=\mathbf{U}\mathbf{\Lambda}\mathbf{U}^{\top}$ we have $(\mathbf{K}_p\mathbf{K}_p^{\top})^{q}\mathbf{K}_p\,\mathbf{\Omega}
      =\mathbf{U}\,\mathbf{\Lambda}^{2q+1}\mathbf{U}^{\top}\mathbf{\Omega}$,
so the leading eigen-directions are amplified by $\lambda_i^{2q+1}$.}.
A thin QR factorisation $\mathbf Y=\mathbf Q\mathbf R$
returns an orthonormal basis
$\mathbf Q\in\mathbb R^{N\times n}$ approximating the column space of  $\mathbf{K}_p$.
The second stage (\emph{eigendecomposition}) applies the Nystr\"{o}m trick
(\cite{halko2011finding}, Alg.~5.5):
\[
\mathbf{B}_1 = \mathbf{K}_p\mathbf Q,
\quad
\mathbf{B}_2 = \mathbf Q^{\top}\mathbf{B}_1,
\]
factor $\mathbf{B}_2$ via a Cholesky decomposition
$\mathbf{B}_2 = \mathbf C^{\top}\mathbf C$, and set $\mathbf{F} = \mathbf{B}_1\,\mathbf C^{-1}$. The \ac{SVD} $\mathbf{F} = \mathbf{U}\,\mathbf{\Sigma}\,\mathbf{V}^{\top}$ with $\mathbf{U}\in\mathbb R^{N\times n}$ and
$\mathbf{\Sigma}\in\mathbb R^{n\times n}$ diagonal, yields the rank-$n$ approximation
$$ \
\mathbf{K}_p \approx
\mathbf{U}\,\mathbf{\Lambda}\,\mathbf{U}^{\top},
\qquad
\mathbf{\Lambda} \coloneqq \mathbf{\Sigma}^{2}
$$
at a cost of $O\bigl((q+1)Nn + n^{3}\bigr)$ flops and $O(Nn)$ memory.
This leads to a preconditioner via the Woodbury method
$$
\mathbf M^{-1} = \bigl(\mathbf{U}\mathbf{\Lambda}\mathbf{U}^{\top} + \eta\mathbf I\bigr)^{-1} = \eta^{-1}\Bigl[ \mathbf{I} - \mathbf{U}\, \bigl(\eta\mathbf{\Lambda}^{-1}+\mathbf{I}\bigr)^{-1}\mathbf{U}^{\top} \Bigr],
$$
which requires only $O(n^{3}+n^{2}N)$ extra work and stores only
$\mathbf{U}$ and $\mathbf{\Lambda}$.

\section{Empirical Assessment}
\label{sec: empirical}

This section presents an empirical comparison of the preconditioners from \Cref{subsec: preconditioners}.
Our test bed is a logistic regression problem described in \Cref{subsec: test bed}. 
The results are presented in \Cref{subsec: results} and the scalability of these methods is discussed in \Cref{subsec: large N}.
Arithmetic was double precision and all computations were implemented in JAX \cite{jax2018github}.
Python code to reproduce these results can be downloaded from \url{https://github.com/MatthewAlexanderFisher/pcg-stein}.

\subsection{Experimental Setup}
\label{subsec: test bed}

\paragraph{Logistic Regression Test Bed:} Logistic regression is a simple example of an analytically intractable posterior distribution for which numerical methods are required.
Full details are reserved for the Online Appendix, but we note that the dimension of the posterior was initially fixed to $d = 4$.
Approximate samples were generated from the posterior distribution obtained using random walk Metropolis--Hastings \ac{MCMC}, and we took the first $N$ \emph{distinct} samples as our collocation nodes $(\mathbf{x}_n)_{1 \leq n \leq N}$.
Initially $N = 10^3$ samples were generated, so that the linear system \eqref{eq: linear-system} could be exactly solved to provide a ground truth against which performance can be measured; performance on tasks with larger $N$ will be discussed in \Cref{subsec: large N}.
For the base kernel $k$ in \eqref{eq: trans invar} we initially used the inverse multi-quadric function $\varphi( \cdot) = (1 + \|\cdot\|^2)^{-1/2}$, which is a common choice for construction of Stein kernels \cite{gorham2017measuring}, and a length scale $\ell > 0$ to be specified.
Through varying the length scale the condition of the matrix $\mathbf{K}_p$ was worsened or improved, so that a range of task difficulties were considered.

\paragraph{Performance Metric:}
For this empirical assessment we focused on a performance metric that is directly related to the motivating numerical task.
Namely, we considered the minimum number of iterations $m_{\text{PCG}}$ of preconditioned \ac{CG} required until the iterate $\mathbf{w}_{m_{\text{PCG}}}$ has worst-case error $\sigma(\mathbf{w}_{m_{\text{PCG}}})$ falling within $1\%$ of the worst-case error $\sigma(\mathbf{w})$ that would be achieved by the exact solution $\mathbf{w}$ of \eqref{eq: linear-system}; i.e. $\sigma(\mathbf{w}_{m_{\text{PCG}}}) < 1.01 \sigma(\mathbf{w})$.
For this assessment, the exact solution $\mathbf{w}$ was approximated using $10^4$ iterations of standard \ac{CG}.
This number $m_{\text{PCG}}$ can be compared to the corresponding number of iterations $m_{\text{CG}}$ required by standard \ac{CG}, and we define the \emph{gain} as 
\begin{align*}
    \mathrm{gain} = \mathrm{ln} \left( \frac{1 + m_{\text{CG}}}{1 + \min\{10^4 , m_{\text{PCG}} \} } \right) . 
\end{align*}
A positive gain indicates an improvement compared to standard \ac{CG}, while a negative gain indicates inferior performance compared to standard \ac{CG}.
Typically $m_{\text{PCG}} \ll 10^4$, but in cases where the $10^4$ iteration limit was reached the algorithm was terminated.  As such, we occasionally under-report negative gain, but positive gains are accurately reported.
This performance measure amounts to focusing on the worst-case performance of the estimator for $f = c + v$ with $v$ constrained to the unit ball of $\mathcal{H}(k_p)$, and avoids the need to focus on a specific test function $f$ in our assessment.
A total of $50$ replicate logistic regression datasets were constructed and the average gain across these different experiments was reported.

Although we do not report timings in these experiments (since these depend heavily on the implementation used), we enforced approximate comparability among the preconditioning methods by insisting that the computational complexity associated with $\mathrm{Act}_{\mathbf{M}^{-1}}$ is $O(N^2)$, so that the overall complexity of preconditioned \ac{CG} is unchanged relative to the standard \ac{CG} method.
For example, in the case of Nystr\"{o}m preconditioning, we take the number $n$ of inducing points to be $n = O(N^{1/2})$, since the computational complexity of $\mathrm{Act}_{\mathbf{M}^{-1}}$ is $O(n^3 + n^2 N)$.
For inversion of the smaller $n \times n$ matrices we used \ac{SVD} with spectral clipping, inflating the smallest singular values to ensure a maximum condition number of $10^9$.

\subsection{Comparison of Preconditioners}
\label{subsec: results}

\begin{figure}[t!]
    \centering
    \includegraphics[width=1\linewidth]{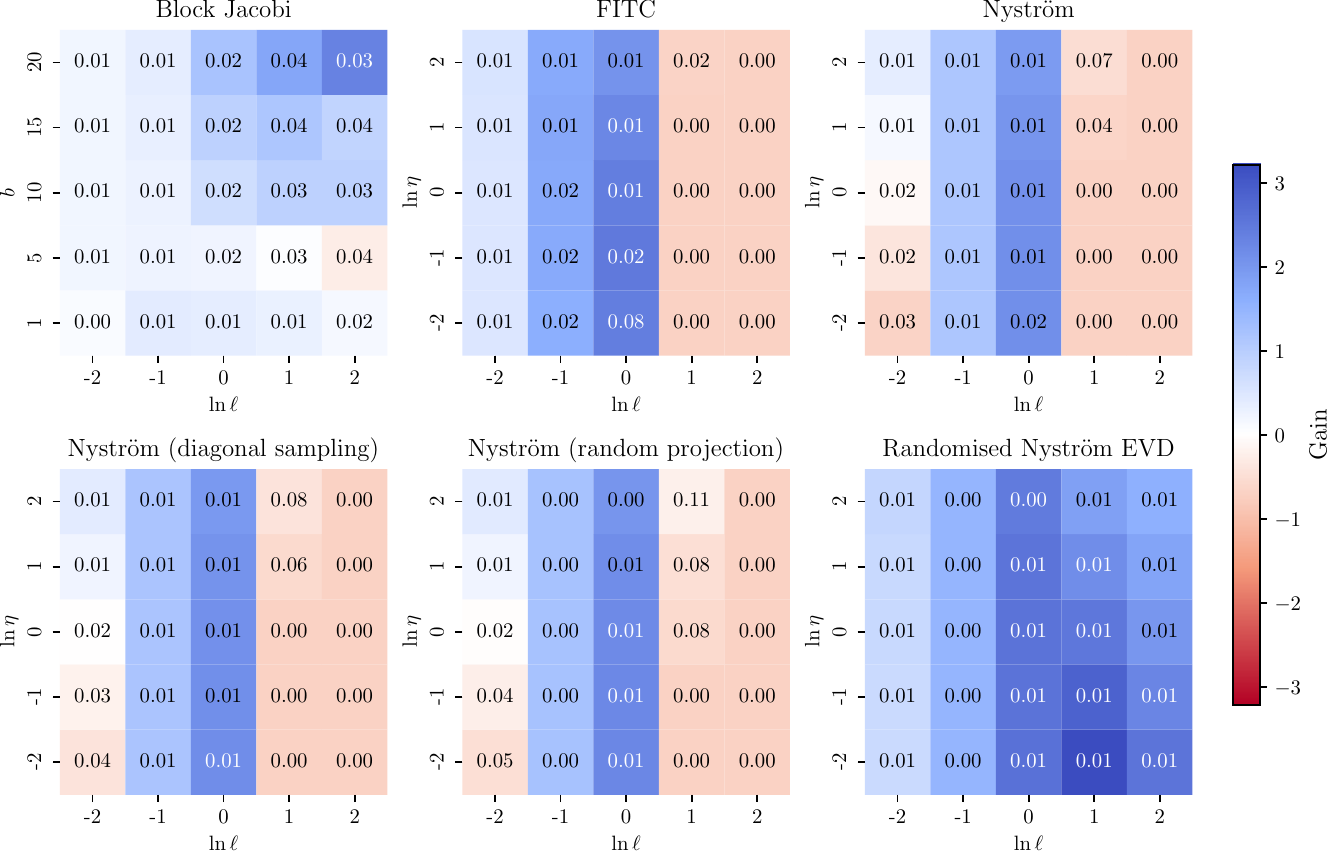}
    \caption{Empirical comparison of preconditioning strategies for fast approximate solution of the canonical Stein equation. 
    (The dimension was $d = 4$.)
    The colours on the heat map show a Monte Carlo estimate of the average gain, calculated over $50$ replicate datasets, while the numbers on the heat map show the standard error associated to each Monte Carlo estimate.
    Blue represents positive gain (improved performance) and red represents negative gain (reduced performance) relative to the standard conjugate gradient method.
    Logarithms of the length scale $\ell$ of the kernel and of the nugget $\eta$ used in Woodbury formula are shown on the horizontal and vertical axes respectively for the second to the sixth preconditioning strategies. 
    For block Jacobi preconditioning, the block size $b$ is shown on the vertical axis.  
    }
\label{fig: main}
\end{figure}

The results of our investigation are summarised in \Cref{fig: main}, where the gain of different preconditioning strategies is displayed as a function of the length scale $\ell$, and as a function of the parameters pertinent to each preconditioning method.
Large values of $\ell$ correspond to matrices that are more ill-conditioned.

First and foremost we note that the randomised Nystr\"{o}m \ac{EVD} performed best over all problem settings considered.
Second, we observe that the gain from all preconditioners was nominal at small values of $\ell$, where the linear system is already well-conditioned.
Third, we observe that all preconditioners except Jacobi deteriorated in performance for large values of $\ell$, with only randomised Nystr\"{o}m \ac{EVD} able to consistently deliver a positive gain in this context.
In favourable settings, gains of 2-3 were observed, corresponding to 10-20 times reduction in the number of iterations required, which is substantial.

\begin{figure}[t!]
\centering
    \includegraphics[width=0.3\linewidth,clip,trim = 13.5cm 0cm 2cm 7.7cm]{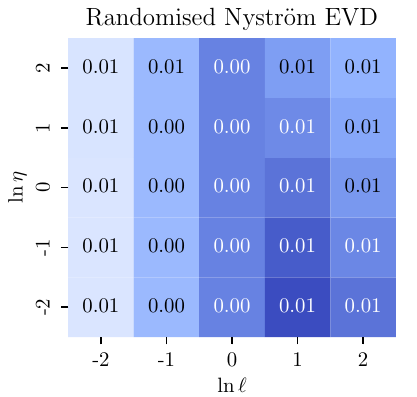}
    \includegraphics[width=0.313\linewidth,clip,trim = 13.5cm 0cm 2cm 7.7cm]{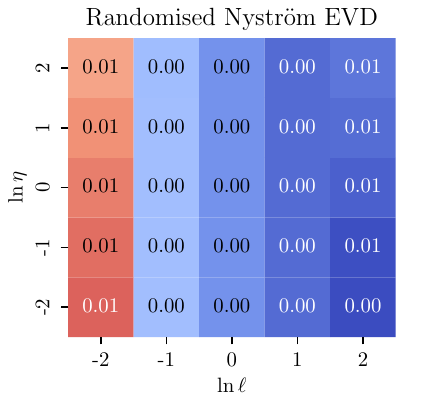}
    \includegraphics[width=0.3\linewidth,clip,trim = 13.5cm 0cm 2cm 7.7cm]{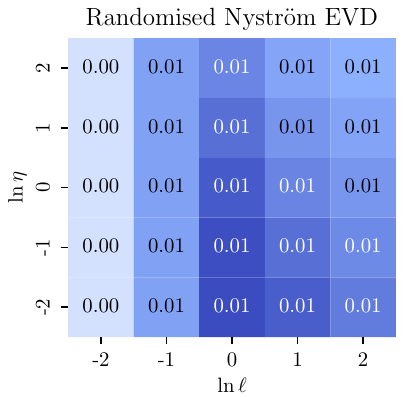}
\caption{Ablation studies for \Cref{fig: main}.
The Mat\'{e}rn $\nu=5/2$ kernel was used in dimension $d = 4$ (left) and dimension $d = 10$ (middle), while the Gaussian kernel was used in dimension $d = 4$ (right).
The same colour scheme as \Cref{fig: main} is used.
Full version in \Cref{fig: main matern52,fig: main 10d,fig: main gaussian} of the Online Appendix.
}
\label{fig: ablation}
\end{figure}

\begin{figure}[t!]
\centering
    \includegraphics[width=1\linewidth,clip,trim = 0.5cm 9.1cm 0cm 0.7cm]{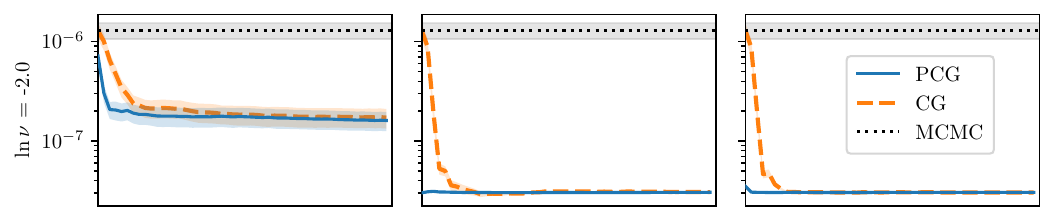}

    \includegraphics[width=1\linewidth,clip,trim = 0.5cm 1cm 0cm 11.5cm]{figures/fig7_evd_1.pdf}
\caption{Squared integration error is reported for a specific integrand (c.f. \eqref{eq: pred inte} in the Online Appendix) as a function of the number of iterations.
The mean squared errors from a total of 50 experiments are reported and standard errors are shaded.
The Gaussian kernel was used in dimension $d = 4$, and the preconditioner was randomised Nystr\"{o}m \ac{EVD}.
As a baseline, the integration error corresponding to an average of the \ac{MCMC} output is presented.
Full version in \Cref{fig: evd1_error} of the Online Appendix.
}
\label{fig: int error summary}
\end{figure}

In the Online Appendix we show that the results we present are not strongly affected by the dimension $d$ of the posterior distribution that defines the numerical task, nor are they strongly affected by the choice of the kernel.
Summarised results are displayed in \Cref{fig: ablation}.
The relevance of these results to integration error for specific integrands is also confirmed in the Online Appendix, with partial results shown in \Cref{fig: int error summary}.

\subsection{Scaling to Large $N$}
\label{subsec: large N}

\begin{figure}[t!]
\centering
\includegraphics[width=0.8\linewidth]{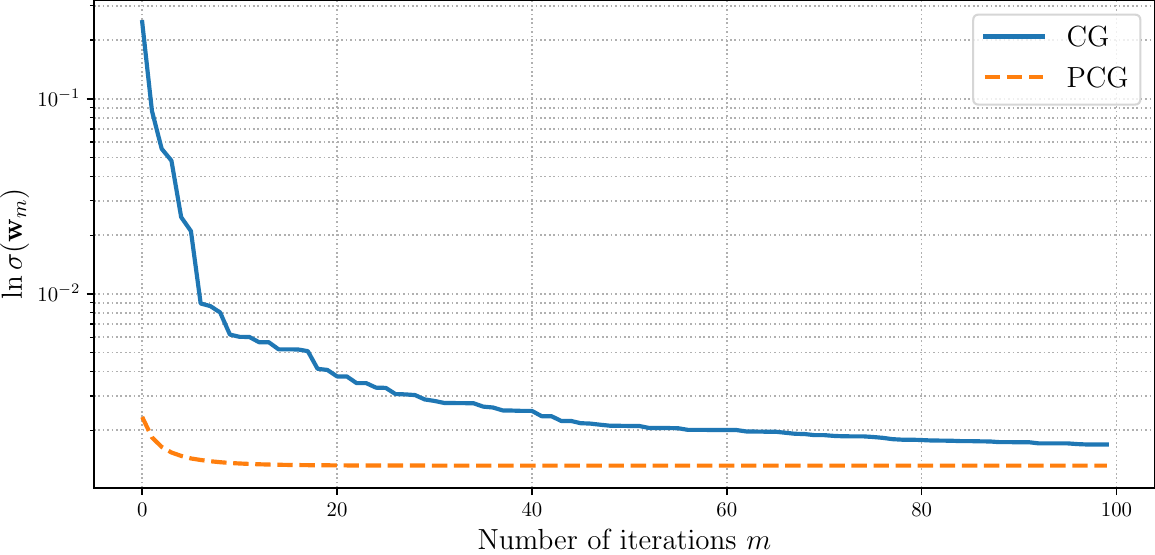}
\caption{Performance of Jacobi preconditioning for large \ac{MCMC} output.
Here the worst-case error $\sigma(\mathbf{w}_m)$ in \eqref{eq: performance bound} was computed as a function of the number $m$ of iterations of both standard and preconditioned \ac{CG}, where the number of \ac{MCMC} iterations resulted in $N = 2 \times 10^4$ distinct collocation nodes.
Here, we set $\ln \ell = 1$ and used the randomised Nyst\"{o}m \ac{EVD} preconditioner with $\ln \eta = -2$ and $n = 200$.
}
\label{fig: large N}
\end{figure}

Though we lose access to a ground truth when $N$ is large, we were curious whether the Jacobi preconditioner might demonstrate gains in this setting, given that it had previously been advocated in the \ac{PDE} context \cite{fasshauer1999solving}.
To investigate, we fixed a logistic regression task and ran $5 \times 10^4$ iterations of \ac{MCMC}, which is a realistic sample size for \ac{MCMC} output, obtaining $N = 20,000$ distinct collocation nodes in total.
The worst-case error $\sigma(\mathbf{w}_m)$ was computed as a function of the number $m$ of iterations for both randomised Nyst\"{o}m \ac{EVD} and standard \ac{CG}.
The results are displayed in \Cref{fig: large N} and show that a non-trivial reduction in approximation error is achieved.

\section{Discussion}
\label{sec: discuss}

Despite considerable recent interest and encouraging proofs-of-concept, the numerical solution of Stein equations remains a challenging task.
This article shed light on the challenge, and explored the potential of numerical preconditioning techniques to improve the cost-accuracy trade-off.
Our main finding was that randomised Nystr\"{o}m \ac{EVD} preconditioning can be an effective strategy in general.
At the same time, we acknowledge the limitations of our empirical investigation, which focused on a particular logistic regression example and samples provided by a particular \ac{MCMC} method.
The cumulative weight of subsequent empirical evidence will be required to determine the generality with which our findings hold.

As a possible avenue for future work, we note that restricting the function space can provide an orthogonal route to reduction in computational cost that can also exploit preconditioning, as exemplified in \cite{rudi2017falkon,meanti2020kernel}.
More speculatively, we note that the same linear system \eqref{eq: linear-system} appears in the \emph{Stein importance sampling} method \cite{liu2017black,hodgkinson2020reproducing,fisher2023gradient,wang2024stein}, albeit in a constrained optimisation context where one wishes to minimise $\mathbf{w}^\top \mathbf{K}_p \mathbf{w}$ subject to $\mathbf{1}^\top \mathbf{w} = 1$ and $\mathbf{w} \geq \mathbf{0}$; we speculate on a useful role for the preconditioning techniques that we have discussed.
Further, in practice one is usually faced with solving a collection of related linear systems (e.g. in exploring the effect of the length scale $\ell$, as we noted in \Cref{subsubsec: kernel choice}).
Techniques such as warm-starting, subspace recycling, and meta-learning may offer a useful additional speed-up in that broader context \cite{parks2006recycling,sun2023meta,lin2024improving}.

\paragraph{Acknowledgments}

QL was supported by the China Scholarship Council 202408060123. 
HK and CJO were supported by EPSRC EP/W019590/1.
CJO was supported by The Alan Turing Institute and a Philip Leverhulme Prize PLP-2023-004. FXB was supported by EPSRC EP/Y022300/1.

\appendix
\section*{Online Appendices}

These appendices are online-only supplement to the manuscript \emph{Fast Approximate Solutions of Stein Equations for Post-Processing of MCMC} by Liu \textit{et al.}, 2025.

\section{Proof of Proposition}

First consider $c \in \mathbb{R}$ to be fixed.
Following identical reasoning to that used in the discussion of collocation methods in the main text, a solution $v_N(\cdot ; c)$ to
\begin{align*}
\argmin_{v \in \mathcal{H}(k_p)} \|v\|_{\mathcal{H}(k_p)} \; \text{such that} \; f(\mathbf{x}_n) = c + v(\mathbf{x}_n) , \qquad n = 1,\dots , N 
\end{align*}
will have the form 
$$
v_N(\mathbf{x} ; c) = \sum_{n=1}^N w_n(c) k_p(\mathbf{x},\mathbf{x}_n),
$$ 
for some weights $\mathbf{w}(c) = [w_1(c),\dots,w_N(c)]^\top$. 
Since
\begin{align*}
    \|v_n(\cdot ; c)\|^2_{\mathcal{H}(k_p)} = \mathbf{w}(c)^\top \mathbf{K}_p \mathbf{w}(c),
\end{align*}
the original optimisation problem is equivalent to
\begin{align*}
    \argmin_{c \in \mathbb{R} , \mathbf{w} \in \mathbb{R}^N } \mathbf{w}^\top \mathbf{K}_p \mathbf{w} \; \text{such that} \; \mathbf{f} = c \mathbf{1} + \mathbf{K}_p \mathbf{w} .
\end{align*}
This is a linearly constrained quadratic programme which can be solved in closed-form using Lagrange multipliers $\bm{\lambda} = [\lambda_1,\dots,\lambda_N]^\top$.
That is, we consider the Lagrangian
\begin{align*}
    \mathcal{L}(c,\mathbf{w},\mathbf{\lambda}) 
        &= \bm{w}^\top \mathbf{K}_p \mathbf{w} - \bm{\lambda}^\top(c\mathbf{1} + \mathbf{K}_p \mathbf{w} - \mathbf{f})
\end{align*}
and solve for the values $c$, $\mathbf{w}$ and $\bm{\lambda}$ for which $\partial_{\bm{\lambda}} \mathcal{L} = \mathbf{0}$, $\partial_{\mathbf{w}} \mathcal{L} = \mathbf{0}$, and $\partial_c \mathcal{L} = 0$.
That is to solve
\begin{align*}
    &c\mathbf{1} + \mathbf{K}_p \mathbf{w} - \mathbf{f} = \mathbf{0} \\
    &2\mathbf{K}_p \mathbf{w} - \mathbf{K}_p \bm{\lambda} = \mathbf{0} \\
    &\bm{\lambda}^\top \mathbf{1}=0.
\end{align*}
Since $k_p$ is a symmetric and positive definite kernel, and since the collocation nodes $(\mathbf{x}_n)_{1 \leq n \leq N}$ are assumed to be distinct, the matrix $\mathbf{K}_p$ is invertible, and we have that $\mathbf{w} = \mathbf{K}_p^{-1}(\mathbf{f}-c\mathbf{1})$ and $\bm{\lambda} = 2 \mathbf{w}$.
Eliminating $\mathbf{w}$ and $\bm{\lambda}$ from the simultaneous equations and solving for $c$, we obtain the stated result.

\section{Pseudocode}

Pseudocode for memory-efficient computation of $\mathrm{Act}_{\mathbf{K}_p}$ is presented in \Cref{mem efficient}.
This pseudocode assumes that the gradients $\mathbf{g}_n := (\nabla \log p)(\mathbf{x}_n)$ have been pre-computed and cached.
The memory bandwidth $B$ should be selected so that a $B \times N$ sub-matrix of $\mathbf{K}_p$ can be comfortably stored in RAM.

\begin{algorithm}[h!]

\algblock{ParFor}{EndParFor}
\algnewcommand\algorithmicparfor{\textbf{parfor}}
\algnewcommand\algorithmicpardo{\textbf{do}}
\algnewcommand\algorithmicendparfor{\textbf{end\ parfor}}
\algrenewtext{ParFor}[1]{\algorithmicparfor\ #1\ \algorithmicpardo}
\algrenewtext{EndParFor}{\algorithmicendparfor}
    
      \caption{Memory-Efficient Multiplication with $\mathbf{K}_p$}
      \begin{algorithmic}[1]
        \Require{$(\mathbf{x}_n)_{1 \leq n \leq N}$ (states), $(\mathbf{g}_n)_{1 \leq n \leq N}$ (gradients), $B$ (memory bandwidth)}
        \Procedure{$\mathrm{Act}_{\mathbf{K}_p}(\mathbf{v})$}{}
        \Let{$\mathbf{a}$}{$\mathbf{0}$} \Comment{initialise $N \times 1$ vector}
        \For{$b = 1, \dots , \lceil N/B \rceil$}  \Comment{for each batch}
            \Let{$(n_{\min},n_{\max})$}{$((b-1)B+1 , \min(N,bB))$}  \Comment{range of data to load}
            \Let{$\bm{\kappa}$}{$\bm{0}$}  \Comment{initialise $(n_{\max}-n_{\min}+1) \times N$ matrix}
            \ParFor{$n_{\min} \leq i \leq n_{\max}$ and $1 \leq j \leq N$}  \Comment{parallel for loop}
            \Let{$\kappa_{i - n_{\min} + 1,j}$}{$\begin{array}{l} (\nabla_1 \cdot \nabla_2 k)(\mathbf{x}_i,\mathbf{x}_j) + (\nabla_1 k)(\mathbf{x}_i,\mathbf{x}_j) \cdot \mathbf{g}_j \\ \hspace{40pt} + \mathbf{g}_i \cdot (\nabla_2 k)(\mathbf{x}_i,\mathbf{x}_j) + k(\mathbf{x}_i,\mathbf{x}_j) \mathbf{g}_i \cdot \mathbf{g}_j \end{array} $}  \Comment{Stein kernel}
            \EndParFor
            \Let{$[a_n]_{n_{\min} \leq n \leq n_{\max}}$}{$\bm{\kappa} [v_n]_{n_{\min} \leq n \leq n_{\max}}$} \Comment{compute $b$th block}
        \EndFor
        \EndProcedure  \Comment{return $\mathbf{a}$}
      \end{algorithmic}
      \label{mem efficient}
    \end{algorithm}

Pseudocode for the preconditioned conjugate gradient method that we implemented in contained in \Cref{the alg}.
The pseudocode does not specify a termination criterion.
For the experiments that we conducted, the termination criterion was $\sigma(\mathbf{w}_m) / \sigma(\mathbf{w})$ and the tolerance was $\tau = 1.01$, meaning that the algorithm terminated when the worst case integration error $\sigma(\mathbf{w}_m)$, defined in \eqref{eq: performance bound}, was within 1\% of the worst case integration error obtained by exactly solving the linear system \eqref{eq: linear-system}.
In practice $\sigma(\mathbf{w})$ will not be exactly known, as this depends on the solution of the linear system  \eqref{eq: linear-system} itself, and an alternative termination criterion will be required; this could be residual-based (e.g. $\Vert \mathbf{r}_m \Vert / \|\mathbf{r}_0\| > \tau$) or simply terminating after a prescribed number of iterations have been performed.

\begin{algorithm}[h!]
      \caption{Preconditioned Conjugate Gradient for the Stein Equation}
      \begin{algorithmic}[1]
        \Require{$\mathrm{Act}_{\mathbf{M}^{-1}}(\cdot)$, $\mathrm{Act}_{\mathbf{K}_p}(\cdot)$, $\mathrm{criterion}(\cdot)$}
        \Procedure{pCG-Stein($\mathbf{w}_0$, $\tau$)}{} \Comment{initial guess $\mathbf{w}_0$ and tolerance $\tau$}
            \Let{$\mathbf{r}_0$}{$\mathbf{1} - \mathrm{Act}_{\mathbf{K}_p}( \mathbf{w}_0) $} \Comment{initial residual}
            \Let{$\mathbf{z}_0$}{$\mathrm{Act}_{\mathbf{M}^{-1}}(\mathbf{r}_0)$}  \Comment{apply inverse preconditioner}
            \Let{$\mathbf{s}_0$}{$\mathbf{z}_0$} \Comment{initial search direction}
            \State $m \gets 0$ \Comment{initialise iteration counter}
            \While{$\mathrm{criterion}(\mathbf{w}_m) > \tau$} \Comment{termination criterion}
            \Let{$\alpha_m$}{$\la \mathbf{r}_m, \mathbf{z}_m\ra / \la \mathbf{s}_m, \mathrm{Act}_{\mathbf{K}_p}( \mathbf{s}_m) \ra$}
            \Let{$\mathbf{w}_{m+1}$}{$\mathbf{w}_m + \alpha_m \mathbf{s}_m$} \Comment{update approximate solution}
            \Let{$\mathbf{r}_{m+1}$}{$\mathbf{r}_m - \alpha_m \mathrm{Act}_{\mathbf{K}_p}( \mathbf{s}_m) $} \Comment{updated residual}
            \Let{$\mathbf{z}_{m+1}$}{$\mathrm{Act}_{\mathbf{M}^{-1}} (\mathbf{r}_{m+1})$}  \Comment{apply inverse preconditioner}
            \Let{$\beta_{m+1}$}{$\la \mathbf{r}_{m+1}, \mathbf{z}_{m+1}\ra / \la \mathbf{r}_m, \mathbf{z}_m\ra$}
            \Let{$\mathbf{s}_{m+1}$}{$\mathbf{z}_{m+1} + \beta_{m+1} \mathbf{s}_m$} \Comment{next search direction}
            \Let{$m$}{$m+1$}  \Comment{update iteration counter}
            \EndWhile
        \EndProcedure  \Comment{return $\mathbf{w}_{m}$}
      \end{algorithmic}
      \label{the alg}
    \end{algorithm}

\section{Logistic Regression Test Bed}
\label{subsec: running}

Logistic regression is perhaps the most commonly encountered example of a Bayesian analysis where the posterior distribution does not admit a closed form, and is therefore a suitable test bed for this work.

\subsection{The Logistic Regression Model}

Suppose we have data $\{(\mathbf{z}_i,y_i)\}_{i=1}^{n_{\text{data}}}$ consisting of \emph{covariates} $\mathbf{z}_i \in \mathbb{R}^d$ and \emph{responses} $y_i \in \{0,1\}$, and the aim is to learn a statistical model capable of predicting responses when covariates are provided.
The \emph{logistic regression} model interprets the responses $y_i$ as realisations of random variables $Y_i$, which are conditionally independent given \emph{parameters} $\mathbf{X} \in \mathbb{R}^d$, with
\begin{equation} \label{eq: logistic-model}
\mathrm{Prob}(Y_i=1| \mathbf{X}) = \rho_i(\mathbf{X}) , \qquad \rho_i(\mathbf{X}) = \frac{1}{1+\exp(-\langle \mathbf{X} , \mathbf{z}_i \rangle) } . 
\end{equation}
It is common to assume that the first element of each vector $\mathbf{z}_i$ is the constant $1$, so that $X_1$ can be viewed as a baseline rate or intercept. 
Conditional on $\mathbf{X}=\mathbf{x}$, the response data admit the probability mass function
$$
p_{\mathbf{Y}|\mathbf{X}}(\mathbf{y}|\mathbf{x})
= \prod_{i=1}^{n_{\text{data}}}\rho_i(\mathbf{x})^{y_i}[1-\rho_i(\mathbf{x})]^{1-y_i} .
$$
The Bayesian framework requires also a prior density function for $\mathbf{X}$, and for this work we took $p_{\mathbf{X}}(\mathbf{x}) \propto \exp(-\|\mathbf{x}\|^2 / 2)$.
In this case the posterior distribution admits a density function
\begin{align}
    p(\mathbf{x}) \equiv p_{\mathbf{X}|\mathbf{Y}}(\mathbf{x}|\mathbf{y}) & \propto p_{\mathbf{X}}(\mathbf{x}) p_{\mathbf{Y}|\mathbf{X}}(\mathbf{y}|\mathbf{x}) \label{eq: logistic-posterior}\\
    & = \exp\left( - \frac{\|\mathbf{x}\|^2}{2} \right) \prod_{i=1}^{n_{\text{data}}}\rho_i(\mathbf{x})^{y_i}[1-\rho_i(\mathbf{x})]^{1-y_i}   \nonumber
\end{align}
for which the normalisation constant $p_{\mathbf{Y}}(\mathbf{y})$ cannot be analytically computed.
As a result, numerical approximation is required to perform Bayesian inference using a logistic regression model, and numerical methods such as \ac{MCMC} are routinely employed.
The logistic regression model is a simple instance of a \emph{generalised linear model}, or more generally of a \emph{statistical model}, and increasing model complexity is usually associated with a more complex posterior distribution, increasing the difficulty of the numerical task.

\subsection{Synthetic Data Generation} \label{app: synth-data-gen}

For the simulation study in the main text reported in \Cref{fig: main}, we set the true data-generating parameter to
$$\mathbf{x}_{\text{True}} = (1, -2, 1, 4)^\top,$$
reflecting a scenario where some covariates are more influential than others.
Then we sampled the covariates $z_i \sim N(\mathbf{0},\mathbf{I}_4)$ independently for $i = 1,\dots,n_{\text{data}}$ with the number of data $n_{\text{data}} = 10^3$ fixed.
Conditional on the covariates and the data-generating parameter $\mathbf{x} = \mathbf{x}_{\text{True}}$, the responses $y_i$ were sampled from the logistic regression model \eqref{eq: logistic-model}.
A single realisation of the dataset was stored before comparing the performance of different preconditioners, so that randomness in the data generation does not act as a confounding factor in the empirical assessment.

\subsection{MCMC Setup and Tuning}

The Metropolis--Hastings algorithm that we used was based on a Gaussian symmetric random walk proposal with covariance $\epsilon^2 \mathbf{I}$, where the variance parameter $\epsilon$ was manually tuned through visual inspection of the trace plot.
The acceptance rate was approximately $0.251$ when $d = 4$, while for $d = 10$ the acceptance rate was approximately $0.059$.
For the experiments reported in the main text, each \ac{MCMC} chain was initialised at the origin and run until 1,000 \emph{distinct} samples were obtained, following a burn-in period of 1,000 iterations.

The full workflow was replicated a total of $50$ times, and average gains (and associated standard errors) were reported.

\section{Experimental Details}

This appendix contains full details required to reproduce the experiments that were reported in the main text.

\subsection{Software and Reproducibility} \label{app: software}

The full package and run instructions are available at:
\begin{center}
\url{https://github.com/MatthewAlexanderFisher/pcg-stein}
\end{center}
Full API and usage guides are online at
\begin{center}
\url{https://matthewalexanderfisher.github.io/pcg-stein/}
\end{center}
Reproducibility is ensured by our  
\begin{itemize}
    \item \textbf{Dependency specification:} The \texttt{pyproject.toml} contained within the repository defines the version number of the dependencies (    \texttt{"jax==0.6.0"}, \texttt{"pyyaml==6.0.2"}). 
    \item \textbf{Experiment configurations:} YAML files in \texttt{experiments/} defining datasets, MCMC settings, and seeds.
\end{itemize}

\subsection{Specific Details for Preconditioners}

\subsubsection{Jacobi Preconditioning}
In our experiment, the sizes of the blocks of preconditioners are $b=1,2,3,4,5$. When the size is $1$, the block Jacobi preconditioner is reduced to be the classic Jacobi preconditioner. If the data size is not divisible by the block size $b$, the size of the last block is the remainder. 

\subsubsection{Nystr\"{o}m Preconditioning}
We used $n=50$ as the number of selected collocation nodes. We used the \texttt{jax.numpy.linalg.pinv} function to calculate the inverse part in the Nystr\"{o}m preconditioner. We use $\eta = 10^{-4}, 10^{-2}, 1, 10^2, 10^4$ as the constant required in the Woodbury inverse formula. The same set of $\eta$ values are used for all the rest preconditioners.

\subsubsection{Nystr\"om Preconditioning and Diagonal Sampling}
We used $n=50$ as the number of selected collocation nodes. 

\subsubsection{Fully Independent Training Conditional}

We selected $n = 50$ inducing points uniformly at random from the $1,000$ data points.

\subsubsection{Randomised Nystr\"om EVD}

We used $n = 50$ as the rank of the approximation. We used \texttt{jax.numpy.linalg.qr} and \texttt{jax.numpy.linalg.svd} function to perform QR and \ac{SVD} decompositions as required.

\section{Additional Experimental Results}

This appendix contains numerical results that concern the choice of kernel, the dimension of the distributional target, and the relevance of our findings to integration error for certain specific choices of integrand.

\subsection{Varying the Kernel}
\label{app: vary kernel}

\begin{figure}[t!]
\centering
    \includegraphics[width=1\linewidth]{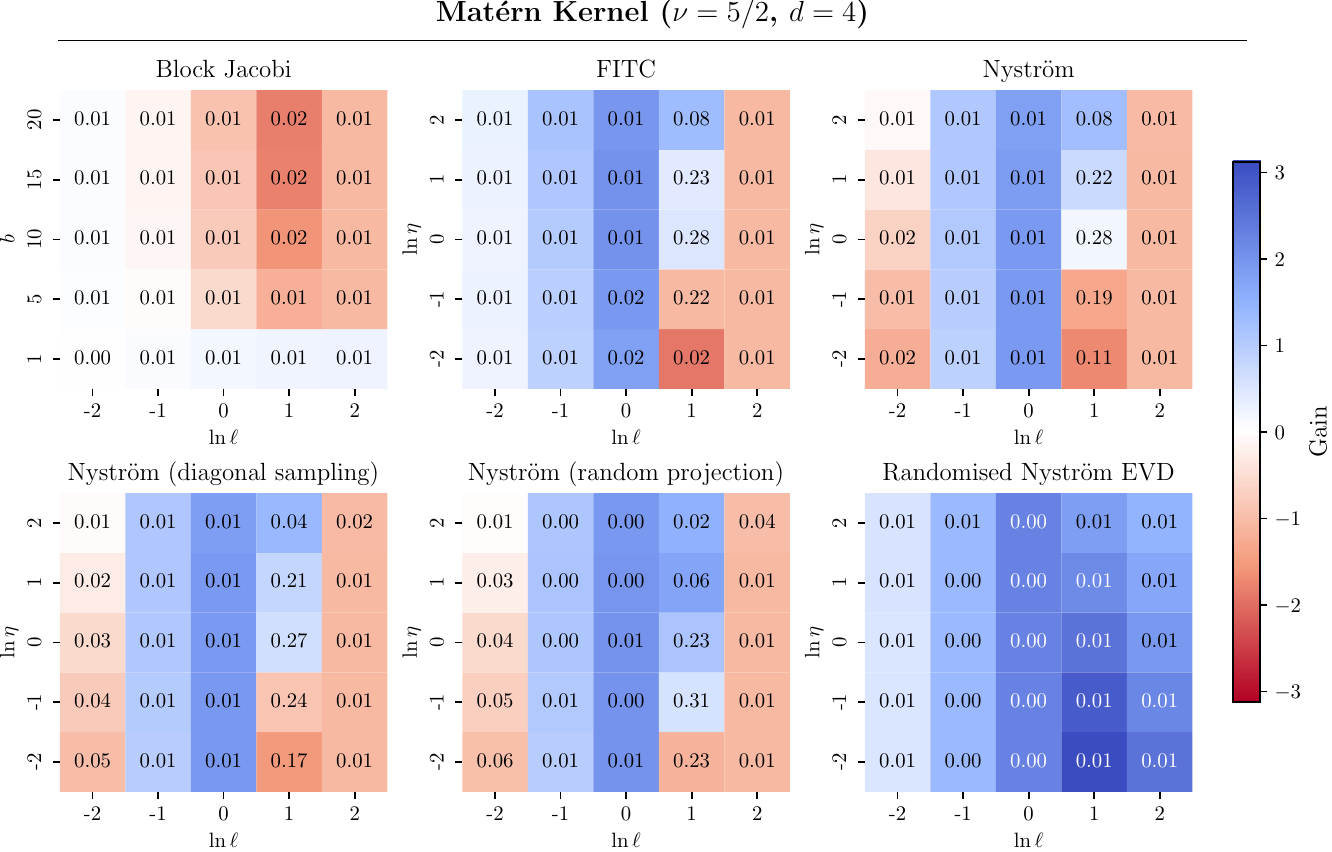}
\caption{Empirical comparison of preconditioning strategies for fast approximate solution of the canonical Stein equation, based on the Mat\'{e}rn $\nu=5/2$ kernel.
}
\label{fig: main matern52}
\end{figure}

\begin{figure}[t!]
\centering
    \includegraphics[width=1\linewidth]{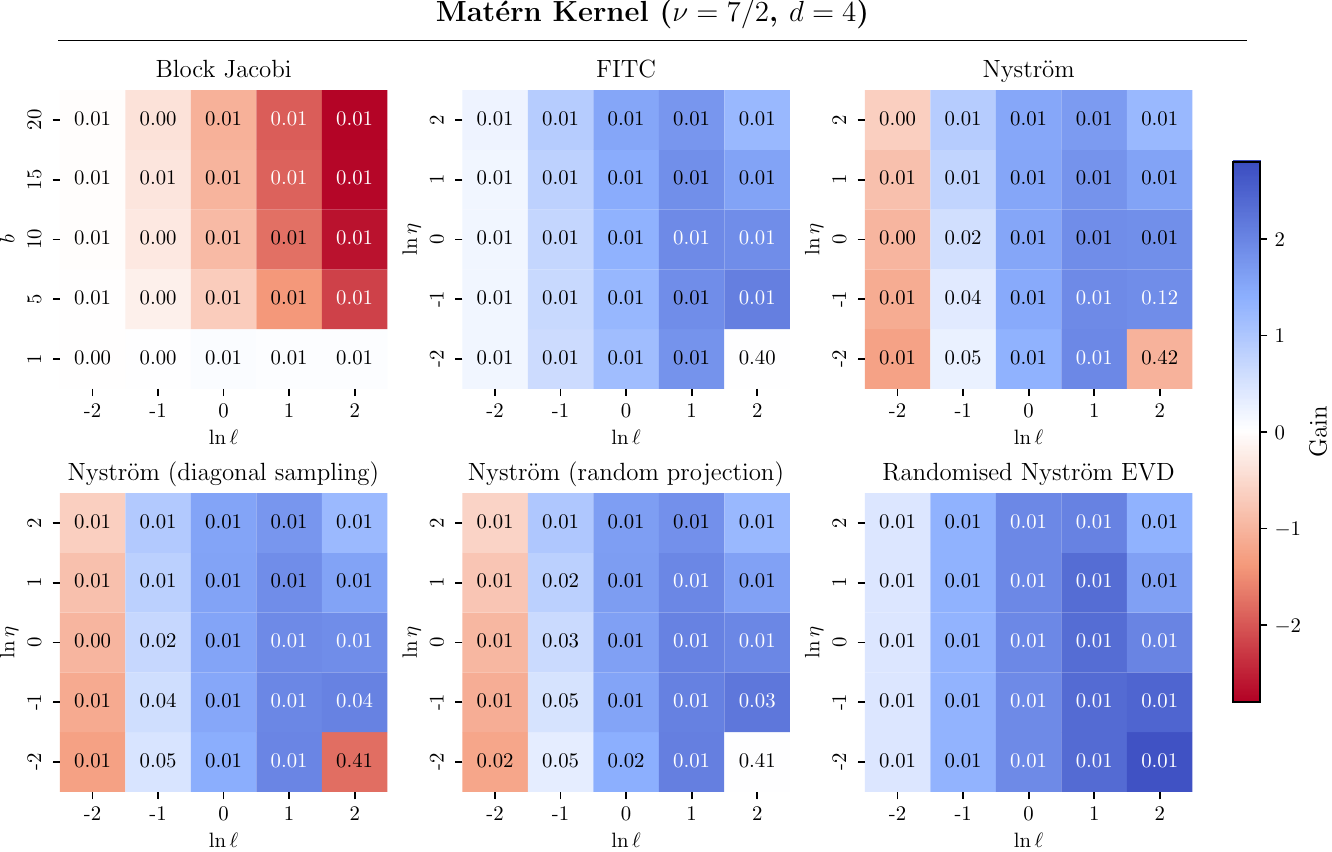}
\caption{Empirical comparison of preconditioning strategies for fast approximate solution of the canonical Stein equation, based on the Mat\'{e}rn $\nu=7/2$ kernel.
}
\label{fig: main matern72}
\end{figure}

\begin{figure}[t!]
\centering
    \includegraphics[width=1\linewidth]{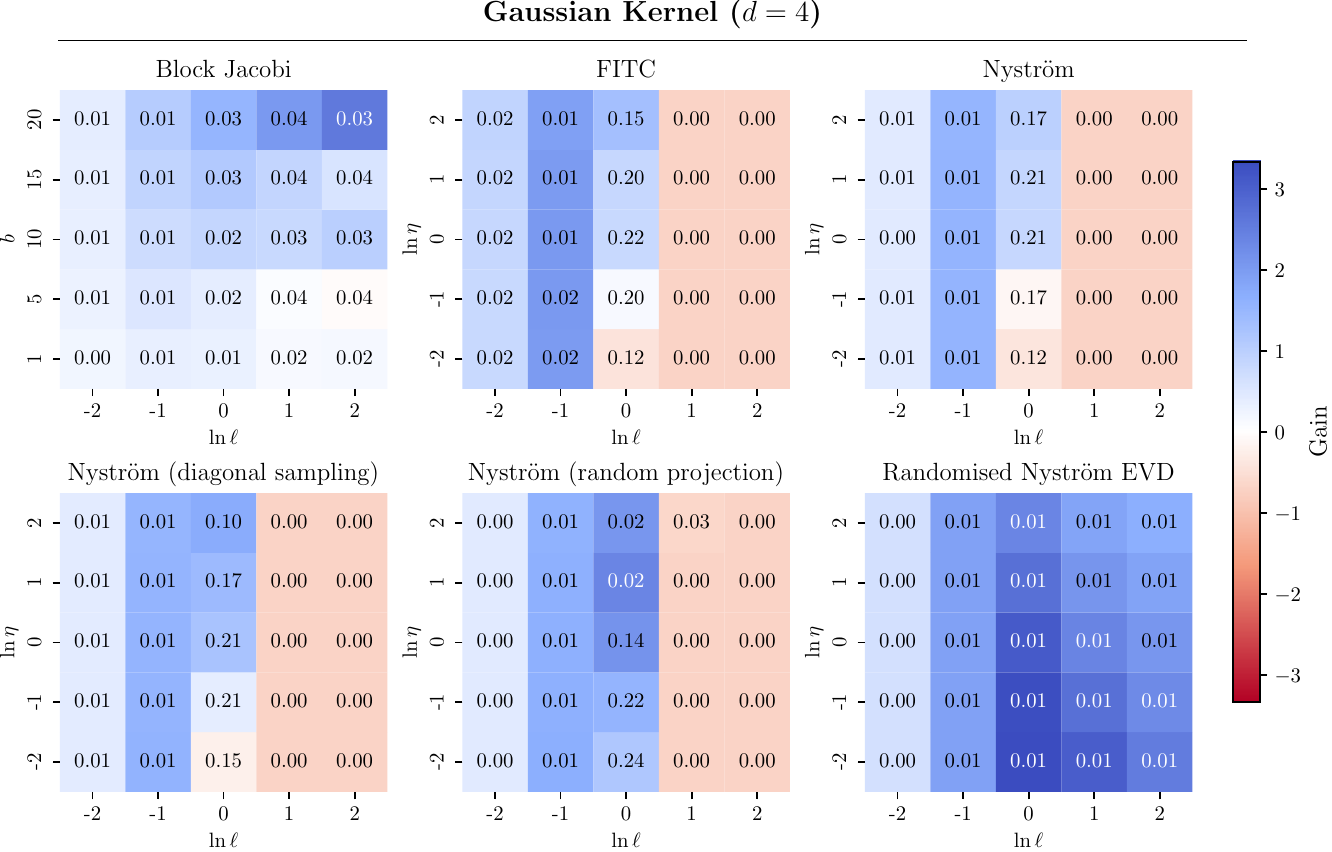}
\caption{
Empirical comparison of preconditioning strategies for fast approximate solution of the canonical Stein equation, based on the Gaussian kernel.
}
\label{fig: main gaussian}
\end{figure}

Here we reproduce the experiment reported in \Cref{fig: main} in the main text, but with different choices of kernel:

\begin{itemize}
     \item \Cref{fig: main matern52}: We use the Mat\'{e}rn kernel $k$ with smoothness parameter $\nu = 5/2$, defined by the radial basis function 
    $$\varphi(r)
    =\sigma^2 \left( 1 + cr + \tfrac{c^2}{3} r^2 \right) e^{-cr},\quad c = \frac{\sqrt{5}}{\ell} .
    $$ 
    Note that $\nu = 5/2$ is the minimum value of $\nu$ for which the Mat\'{e}rn kernel can be explicitly computed and for which the kernel possesses the level of differentiability required.
    \item \Cref{fig: main matern72}: We use the Mat\'{e}rn kernel $k$ with smoothness parameter $\nu = 7/2$, defined by the radial basis function 
    $$\varphi(r)
    =\sigma^2 \Bigl(1+c r+\tfrac{c^{2}}{3}r^{2}+\tfrac{c^{3}}{15}r^{3}\Bigr) e^{-c r},
    \quad
    c=\frac{\sqrt7}{\ell} .
    $$ 
    \item \Cref{fig: main gaussian}: We use the Gaussian kernel $k$, defined by the radial basis function
    $$ \varphi(r) = \sigma^2\exp\left(-\frac{r^2}{2\ell^2}\right) .$$
\end{itemize}

\noindent Compared to \Cref{fig: main} in the main text, performance of the Jacobi preconditioner collapsed when either of the Mat\'{e}rn kernels were used, while the other preconditioners performed broadly as described in the main text.

\subsection{Varying Dimension}
\label{app: vary dim}

\begin{figure}[t!]
\centering
    \includegraphics[width=1\linewidth]{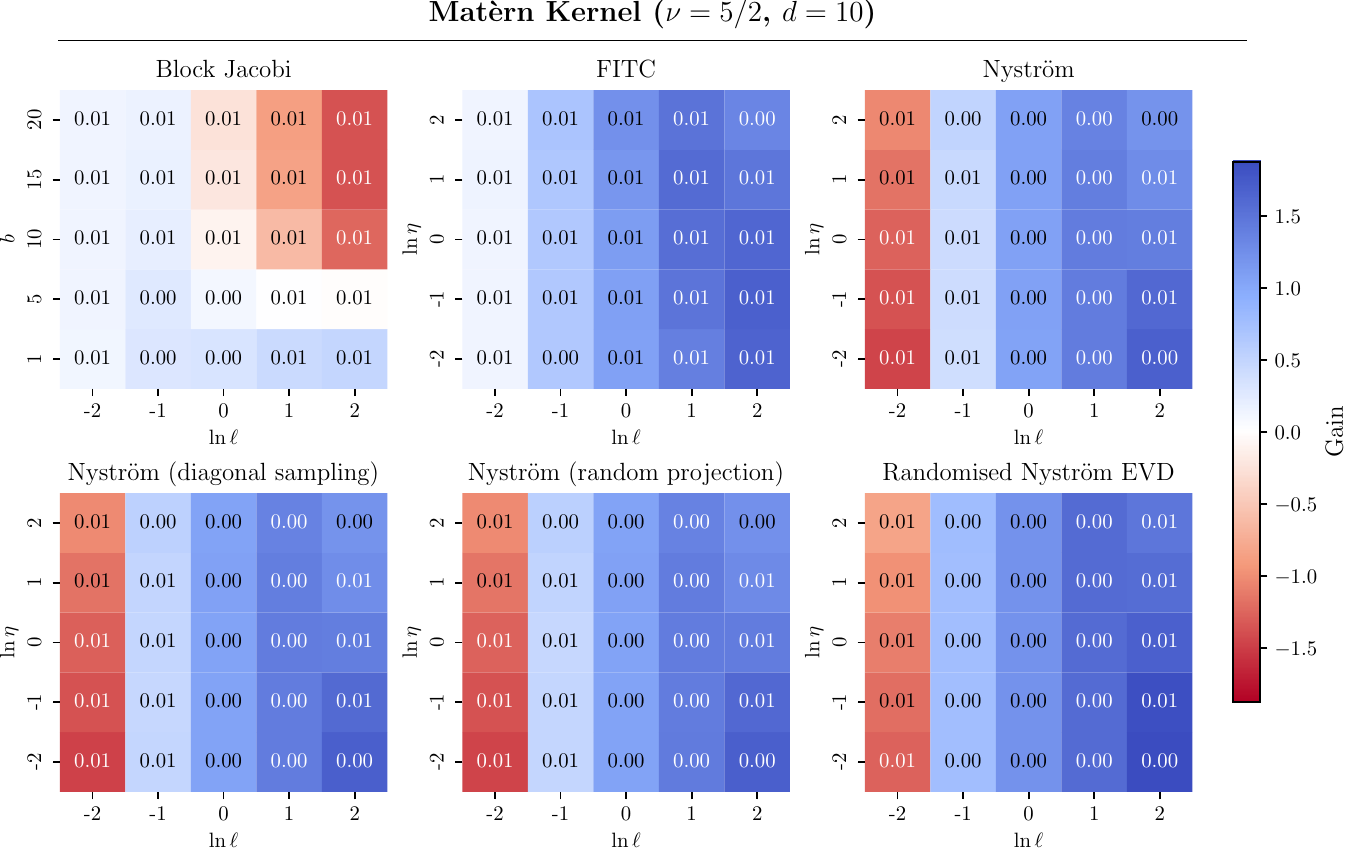}
\caption{Empirical comparison of preconditioning strategies for fast approximate solution of the canonical Stein equation.
}
\label{fig: main 10d}
\end{figure}

Here, we vary the dimension $d$ of the logistic regression task from $d = 4$ in the main text to $d = 10$, to understand how the performance of different preconditioners depends on the dimension of the posterior distribution defining the numerical task.
We set the true data-generating parameter to 
$$\mathbf{x}_{\text{True}} = (1,0.9,0.8, \ldots, 0.1)^\top.$$
Results are reported for the Mat\'{e}rn kernel with $\nu = 5/2$ in \Cref{fig: main 10d}.
The performance of preconditioners was broadly similar to the case $d = 4$ presented in the main text, with the exception that some preconditioners now exhibited negative gain for small $\ell$ at which the linear system is well-conditioned.

\subsection{Integration Error}
\label{app: integration error}

\begin{figure}[t!]
\centering
    \includegraphics[width=1\linewidth]{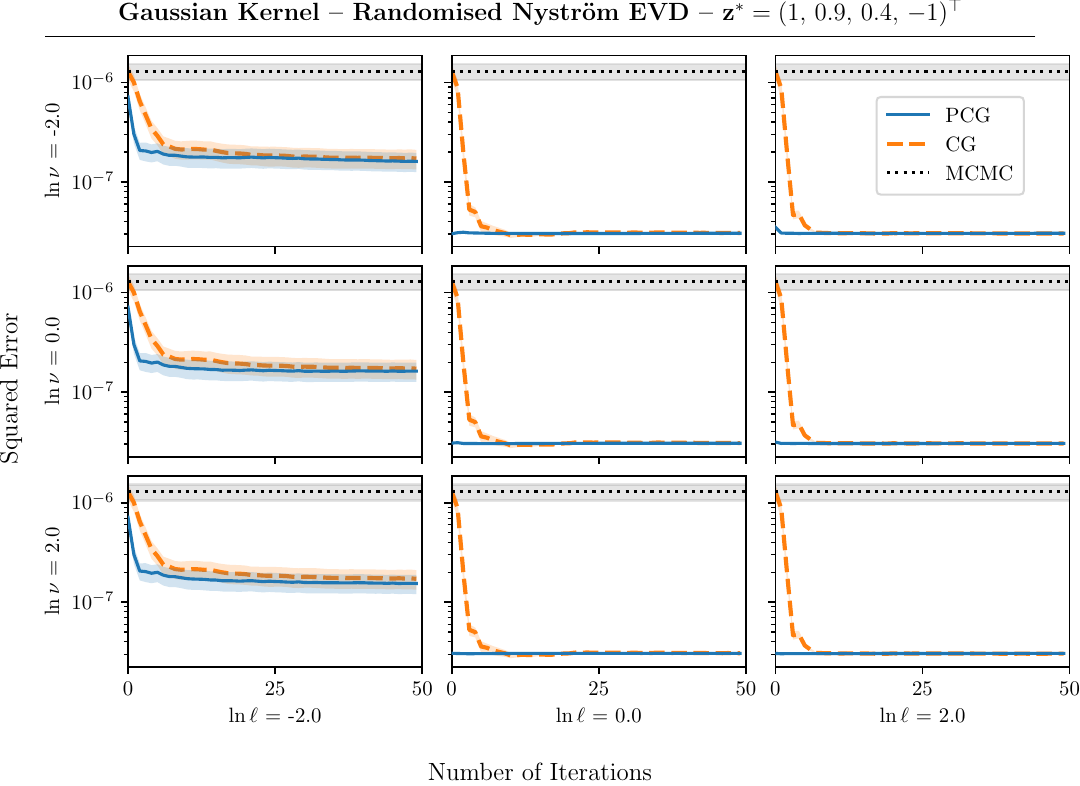}
\caption{Empirical comparison of preconditioning strategies for fast approximate solution of the canonical Stein equation.  Squared integration error is reported for the integrand \eqref{eq: pred inte} as a function of the number of iterations.
The mean squared errors from a total of 50 experiments are reported and standard errors are shaded.
As a baseline, the integration error corresponding to an average of the \ac{MCMC} output is presented.
}
\label{fig: evd1_error}
\end{figure}

\begin{figure}[t!]
\centering
    \includegraphics[width=1\linewidth]{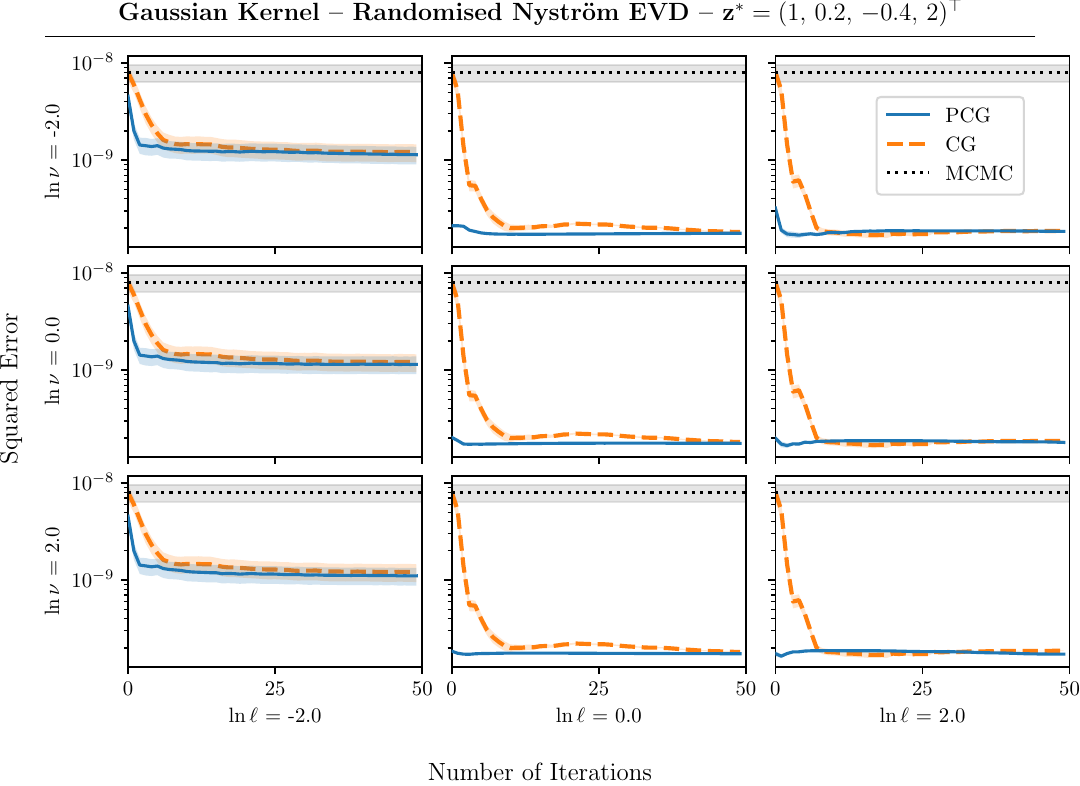}
\caption{Empirical comparison of preconditioning strategies for fast approximate solution of the canonical Stein equation. Squared integration error is reported for the integrand \eqref{eq: pred inte} as a function of the number of iterations.
The mean squared errors from a total of 50 experiments are reported and standard errors are shaded.
As a baseline, the integration error corresponding to an average of the \ac{MCMC} output is presented.
}
\label{fig: evd2_error}
\end{figure}

\begin{figure}[t!]
\centering
    \includegraphics[width=1\linewidth]{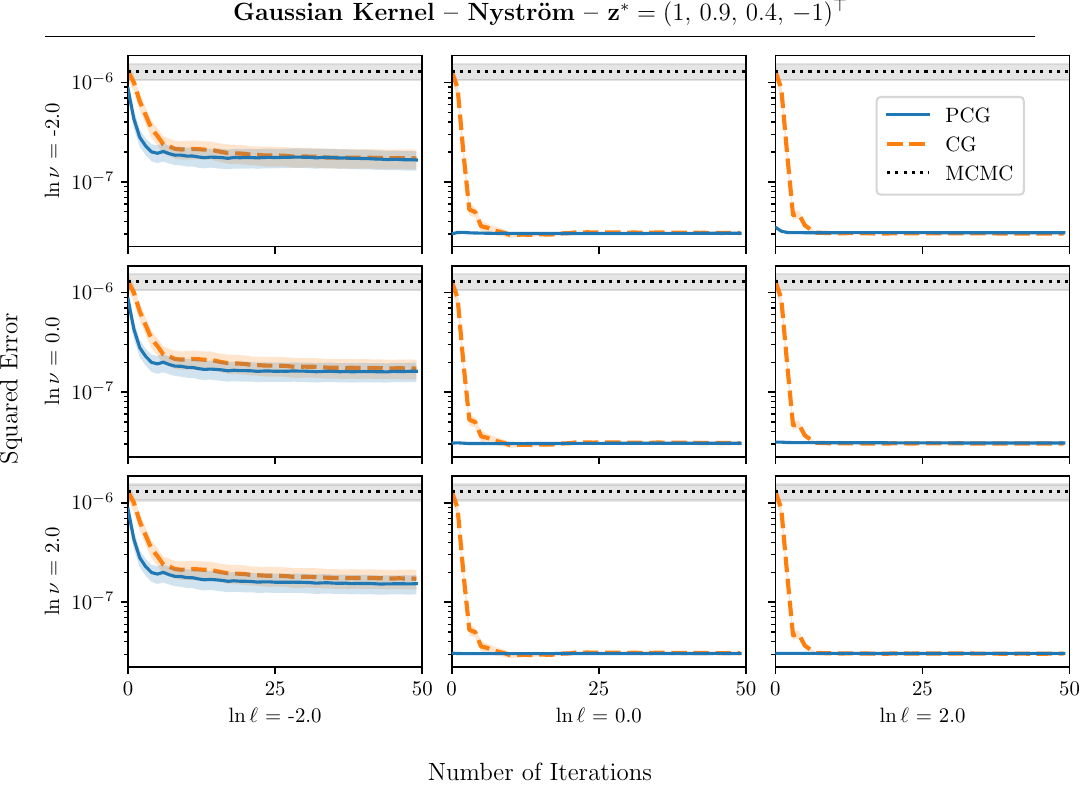}
\caption{Empirical comparison of preconditioning strategies for fast approximate solution of the canonical Stein equation. Squared integration error is reported for the integrand \eqref{eq: pred inte} as a function of the number of iterations.
The mean squared errors from a total of 50 experiments are reported and standard errors are shaded.
As a baseline, the integration error corresponding to an average of the \ac{MCMC} output is presented.
}
\label{fig: nystrom1_error}
\end{figure}

\begin{figure}[t!]
\centering
    \includegraphics[width=1\linewidth]{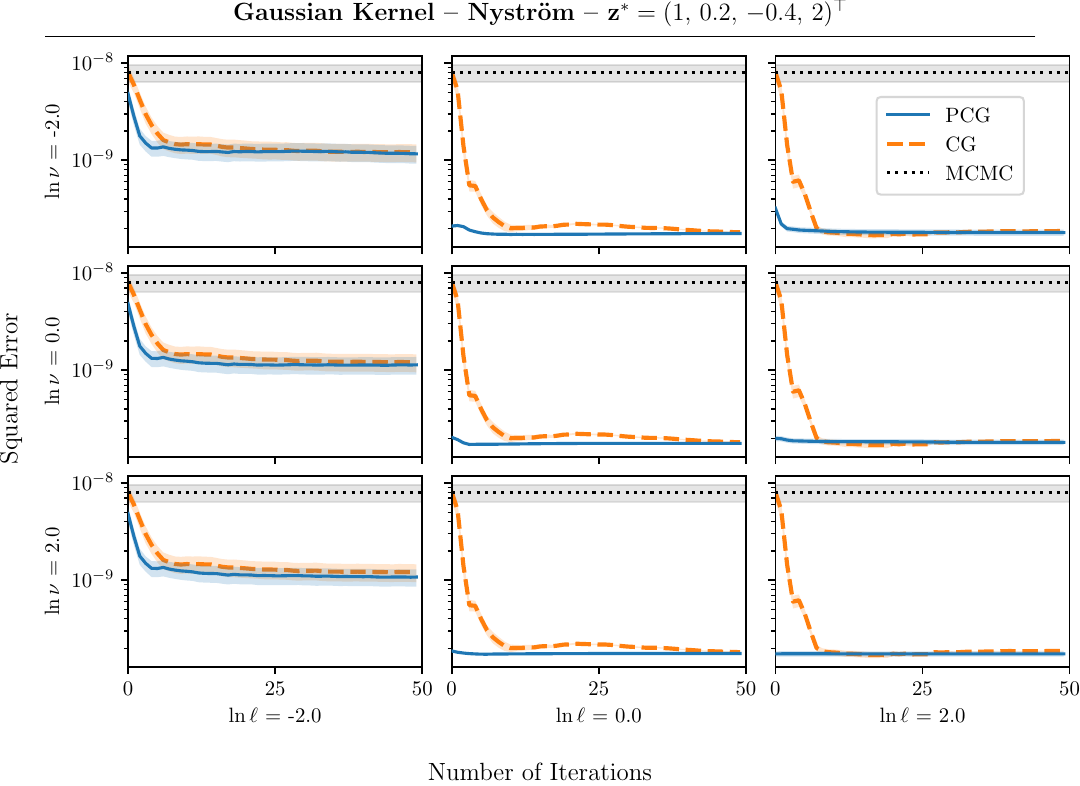}
\caption{Empirical comparison of preconditioning strategies for fast approximate solution of the canonical Stein equation. Squared integration error is reported for the integrand \eqref{eq: pred inte} as a function of the number of iterations.
The mean squared errors from a total of 50 experiments are reported and standard errors are shaded.
As a baseline, the integration error corresponding to an average of the \ac{MCMC} output is presented.
}
\label{fig: nystrom2_error}
\end{figure}

Here we reproduced the experiments reported in the main text, but reporting the absolute integration error for posterior predictives
\begin{align}
f_i(\mathbf{x}) = \int \rho^*_i(\mathbf{x}) \, p(\mathbf{x}) \; \mathrm{d}\mathbf{x}, \label{eq: pred inte}
\end{align}
where $p(\mathbf{x}) = p_{\mathbf{X}\mid\mathbf{Y}}(\mathbf{x} \mid \mathbf{y} )$ is the posterior \eqref{eq: logistic-posterior} and $i = 1, 2$ with
\begin{align*}
    \mathbf{\rho}_i^*(\mathbf{x}) &= \frac{1}{1 + \exp\left(-\langle \mathbf{x}, \mathbf{z}^*_i\rangle\right)} , \qquad
    \begin{array}{ll} \mathbf{z}^*_1 &= (1,0.9, 0.4,-1)^\top \\
    \mathbf{z}^*_2 &= (1,0.2,-0.4, 2)^\top \end{array} 
\end{align*}
instead of the worst-case integration error we report in the main text.
Note that such functions do \emph{not} necessarily belong to the Hilbert space reproduced by the Stein kernel \cite{kanagawa2022controlling}.
As a baseline, the true integral was capproximated using $N = 10^5$ samples generated from \ac{MCMC}. 
Results are presented in \Cref{fig: evd1_error,fig: evd2_error,fig: nystrom1_error,fig: nystrom2_error}.
It can be seen that in all cases the estimator $c_N$ in \eqref{eq: cN} provides a more accurate approximation to the true integral compared to the standard \ac{MCMC} estimator \eqref{eq: mcmc estimator}.
Further, the use of preconditioning accelerates the convergence of the squared integration error in all cases considered.
These results are consistent with the results for the worst-case error metric $\sigma(\cdot)$ which we focused on in the main text.

\bibliographystyle{spmpsci}
\bibliography{bibliography}

\end{document}